\documentclass[aps,prd,preprint,superscriptaddress,nofootinbib]{revtex4-1}

\usepackage{amssymb}
\usepackage{amsmath}
\usepackage{graphicx}
\usepackage{subfigure}
\usepackage{longtable}
\usepackage{color}
\usepackage{textcomp}
\usepackage{hyperref}

\begin{document}

\title{Mechanical loss in state-of-the-art amorphous optical coatings}

\date{\today}

\author{Massimo Granata}
	\email[]{m.granata@lma.in2p3.fr}
\author{Emeline Saracco}
	\altaffiliation[Current affiliation: ]{ISORG, 60 rue des Berges, Parc Polytec, Immeuble Tramontane, 38000 Grenoble, France}
\author{Nazario Morgado}
	\altaffiliation[Current affiliation: ]{Laboratoire de tribologie et de dynamique des syst\`emes, Ecole Centrale de Lyon, CNRS (UMR 5513), 36 avenue Guy de Collongue, 69134 Ecully, France}	
	\affiliation{Laboratoire des Mat\'eriaux Avanc\'es, CNRS (USR 3264), Universit\'e de Lyon, F-69622 Villeurbanne, France}
\author{Alix Cajgfinger}
	\affiliation{Laboratoire des Mat\'eriaux Avanc\'es, CNRS (USR 3264), Universit\'e de Lyon, F-69622 Villeurbanne, France}
\author{Gianpietro Cagnoli}
	\affiliation{Laboratoire des Mat\'eriaux Avanc\'es, CNRS (USR 3264), Universit\'e de Lyon, F-69622 Villeurbanne, France}
	\affiliation{Institut Lumi\`ere Mati\`ere, CNRS (UMR 5306), Universit\'e de Lyon, F-69622 Villeurbanne, France}
\author{J\'{e}r\^{o}me Degallaix}
\author{Vincent Dolique}
\author{Dani\`{e}le Forest}
\author{Janyce Franc}
	\altaffiliation[Current affiliation: ]{Claranor, Chemin de la Rollande Agroparc, BP 21531, 84916 Avignon cedex 9, France}
\author{Christophe Michel}
\author{Laurent Pinard}
	\affiliation{Laboratoire des Mat\'eriaux Avanc\'es, CNRS (USR 3264), Universit\'e de Lyon, F-69622 Villeurbanne, France}
\author{Raffaele Flaminio}
	\affiliation{Laboratoire des Mat\'eriaux Avanc\'es, CNRS (USR 3264), Universit\'e de Lyon, F-69622 Villeurbanne, France}
	\affiliation{National Astronomical Observatory of Japan, 2-21-1 Osawa, Mitaka, 181-8588 Tokyo, Japan}
	
\begin{abstract}
We present the results of mechanical characterizations of many different high-quality optical coatings made of ion-beam-sputtered titania-doped tantala and silica, developed originally for interferometric gravitational-wave detectors. Our data show that in multi-layer stacks (like high-reflection Bragg mirrors, for example) the measured coating dissipation is systematically higher than the expectation and is correlated with the stress condition in the sample. This has a particular relevance for the noise budget of current {\it advanced} gravitational-wave interferometers, and, more generally, for any experiment involving thermal-noise-limited optical cavities.
\end{abstract}

\pacs{xx.yy.zz, ii.jj.ll}

\keywords{any}

\maketitle

\section{Introduction}
The experimental research devoted to the detection of gravitational waves is nowadays entering a critical phase. After having accomplished a first stage of successful operation, the largest ground-based interferometric detectors are presently undergoing a major technical upgrade towards a tenfold sensitivity improvement over the whole detection band \cite{AdVirgo, aLIGO}. Hopefully, this will lead to the first signal detection within the next few years, thus triggering the beginning of gravitational-wave astronomy.

In those detectors, laser interferometry is used to measure the relative displacement of large, massive suspended mirrors playing the role of gravitational-field probes: the test masses \cite{Weiss72, Adhikari14}. In the central and most sensitive region of the detection band, around 200 Hz, the Advanced Virgo and Advanced LIGO interferometers will be limited by the thermal noise of the mirror high-reflection coatings. Also the future Einstein Telescope \cite{ET}, planned to be one order of magnitude more sensitive than advanced detectors, will be affected by this noise source as well. 

The reduction of coating thermal noise is of fundamental importance. For gravitational-wave interferometers, lowering such noise would yield an increase of the detection horizon (i.e. the radius of the accessible volume of the Universe) and hence of the expected detection rate. More generally, lowering the coating thermal noise would be highly beneficial also for other precision experiments using high-finesse optical cavities, such as frequency standards for laser stabilization \cite{Kessler12}, quantum computing devices \cite{Martinis05} and optomechanical resonators \cite{Kippenberg07}.

Thermal noise arises from fluctuations of the mirror surface under the random motion of particles in coatings and substrates \cite{Saulson90, Levin98}. Its amplitude is determined by the amount of internal friction within the mirror materials, via the fluctuation-dissipation theorem \cite{CallenGreene52}: the higher the loss, the higher the thermal noise level. At the same time, since the coating loss is usually several orders of magnitude larger than the substrate one, the thermal noise of the thin high-reflection coating turns out to be largely dominant over that of the thick underlying substrate. Therefore, since the first works that addressed this relevant source of noise \cite{Crooks02, Harry02}, a considerable research effort has been committed to the investigation and the reduction of the mechanical loss of optical coatings\footnote{Assuming a given temperature and laser spot size. Otherwise coating thermal noise could be attenuated of course by lowering the temperature of the optics \cite{Somiya12}, or by increasing the laser spot size and even by using non-Gaussian laser beams \cite{Tarallo07, Bondarescu08, Gatto14}.}.

The high-reflection coatings of gravitational-wave interferometers are Bragg reflectors composed of alternate layers of ion-beam-sputtered (IBS) silica (SiO$_2$) and tantala (Ta$_2$O$_5$), as low- and high-refractive-index materials, respectively. The historical reason for this choice is the low optical absorption of tantala at the typical wavelength of operation of the detectors ($\lambda_0=1064$ nm), as the use of this material allowed reducing the coating absorption down to about 1 ppm \cite{Beauville04}. From the point of view of the mechanics, however, tantala proved to be substantially more dissipative than silica \cite{Penn03, Crooks04}, making it the dominant source of loss in the coating stack.

Eventually, the coating loss was remarkably improved by applying a titania (TiO$_2$) doping to tantala \cite{Harry06, Harry07}. This procedure, developed by the Laboratoire des Mat\'eriaux Avanc\'es (LMA) in Lyon, significantly decreased the mechanical loss of the high-index material of the stack. Moreover, it proved to be beneficial to its optical performances as well \cite{Flaminio10}. Very recently, for example, we measured an average absorption of 0.14 $\pm$ 0.05 ppm on the central area ($\varnothing\,$ 160 mm) of a coating deposited on a test mass of Advanced LIGO. In the meantime, an optimization technique to dilute the loss contribution of the high-index material has also been developed \cite{Agresti06}, further decreasing the resulting coating thermal noise \cite{Villar10}.

In this paper, we will present the results of mechanical characterizations of state-of-the-art coatings of titania-doped tantala and silica, carried out at the LMA since 2009. This activity is part of a wide series of follow-up quality checks (surface quality, absorption and scattering among others) of the optics manufactured at the LMA, especially for advanced gravitational-wave interferometers. Within this frame, we measured the mechanical loss and the internal residual stress of a large series of samples with different features. Coatings of assorted thickness (ranging from few hundreds of nanometers to several microns) and morphology have been characterized on substrates of different geometries. Because of the large number and variety of the samples, our results constitute an important database and set the basis for further research and development of coatings of high optical quality and low mechanical loss. Besides, our measurements reveal the presence of an excess loss whose origin is still not clearly understood. A possible interpretation of this phenomenon will be addressed, and the consequences of our results will be discussed.

\section{Experiment}
In order to characterize the intrinsic loss of several high-reflection optical coatings, we applied the resonance method \cite{NowickBerry72} on a large set of different mechanical resonators. For each of these samples, we measured the ring-down time $\tau$ of several vibrational modes, before and after the deposition of a coating stack. The calculation of the coating mechanical loss then follows straightforwardly from the measurements, provided that the the \textit{dilution factor} $D$ -- the ratio of the elastic energy of the coating to the elastic energy of the coated substrate -- is known \cite{Harry02, Li14}. This coefficient is critical since it accounts for the fact that the energy loss of the coated resonator is diluted by the relatively lossless substrate. To compute this dilution factor, we performed a mechanical finite-element simulation of the coated samples. We also used these simulations to identify the vibrational modes measured in our setups.

In parallel, we used an interferometric microscope to measure the curvature induced by coating internal stress in the coated samples. By associating each value of loss to a radius of curvature, we could study a potential correlation with mechanical loss. To determine the coating morphology, we also carried out a series of analyses with a transmission electron microscope.  

A detailed description of our samples, experimental setups, measurement techniques and simulations is reported in the following. The results of the measurements are presented in Section \ref{SECexpRes}.

\subsection{Samples}
The large majority of our samples is constituted by fused-silica cantilever blades provided by Heraeus\footnote{\url{heraeus-quarzglas.com}} and optically polished by Gooch and Housego\footnote{\url{goochandhousego.com}} and by HTM\footnote{\url{quartz-htm.com}}, having the following dimensions: 5 mm x 42 mm, for a nominal thickness of 110 $\mu$m. High-purity fused silica (Suprasil 311) was chosen because of its very low loss at room temperature \cite{Ageev04}, which makes it very well suited for coating characterization. These cantilevers are welded on a corner to a cylindrical clamping block of the same material, used to suspend the sample. The welding has been performed by hand, with a hydrogen torch, by WisePower\footnote{\url{wisepower.it}}.

The other few substrates are fused-silica disks ($\varnothing\,$ 3") of Corning 7980\footnote{\url{corning.com}} of a nominal thickness of 2.54 mm, micro-polished ($\lambda/10$ flatness, roughness $<0.5$ nm rms) on both surfaces by General Optics\footnote{Now a division of Gooch and Housego}. The purpose of such high-quality polishing is to test the coatings on substrates with specifications as close as possible to those of substrates of gravitational-wave interferometers.
  
The high-reflection coatings are stacks of titania-doped tantala (TiO$_2$Ta$_2$O$_5$) and silica for operation at $\lambda_{0}=1064$ nm (except for only two samples designed for $\lambda$ = 1529 nm). In the doped-tantala layers, the atomic ratio of Ti to Ta is approximately 1/3, as measured through Rutherford backscattering spectrometry. The thickness of each layer is either one-quarter of the laser wavelength (i. e. the typical quarter-wave coating) or optimized \cite{Villar10}. In the latter case, the thickness of each doped-tantala layer is reduced to decrease the loss of the stack and thus its expected thermal noise amplitude, but the overall optical performances are preserved. Nonetheless, even in the optimized configuration, the coating stack consists in a periodic structure of doped tantala and silica pairs, or ‘doublets’, with the exception of the terminal layers (the bottom one is of doped tantala and slightly thicker, while the top one is the thinnest of the stack and is of silica for protective purposes). One example of such coating is the one developed by the LMA for the end test masses of the over-coupled Fabry-Perot arm cavities of Advanced LIGO and Advanced Virgo \cite{Granata13}: it features a measured transmissivity of $\text{T}<5$ ppm and a measured absorption of $<0.5$ ppm, having at the same time nearly the minimum expected thermal noise achievable for this given high reflectivity.

A small subset of substrates are also coated with anti-reflection stacks, for operation at the same $\lambda_{0}$. In such coatings, whose transmissivity is highly sensitive to thickness errors, the number $N$ of alternating layers is reduced to the minimum (usually $2\leq N \leq 6$), and the design of each layer thickness is only driven by the desired optical performances.

Another subset of substrates is coated with a quarter-wave single layer (\textit{mono-layer} coating) either of titania-doped tantala or silica of varying thickness, in order to measure the intrinsic loss of each coating material composing the stacks.

Finally, some substrates have been coated on both surfaces (either with stacks or with mono-layers) in order to change the resulting internal stress of the coated samples themselves. The complete list of the samples used in this work can be found in the tables of Section \ref{SECexpRes}.

All the coatings presented here have been deposited at the LMA through IBS with the same deposition parameters (source settings, substrate temperature, etc.), into a large ($V=10$ m$^3$) custom-developed coating chamber. This chamber has been built specifically to coat twin pairs of large mirrors (each one having $\varnothing\,$ 350 mm and being 200 mm thick) for gravitational-wave interferometers, with very high thickness uniformity over a large surface \cite{Michel12}. As part of the standard post-deposition process, all the coated samples underwent in-air annealing at $T_{A}=500{^\text{ o}}$C over 10 hours. This is necessary to decrease both the internal stress and the optical absorption of the coatings. All the measurements of curvature and mechanical loss presented hereafter have been carried out on annealed samples.

\subsection{Setups}

\subsubsection{Radius of curvature (RoC)}
\label{SECroc}
\begin{figure}
	\includegraphics[width=0.50\textwidth]{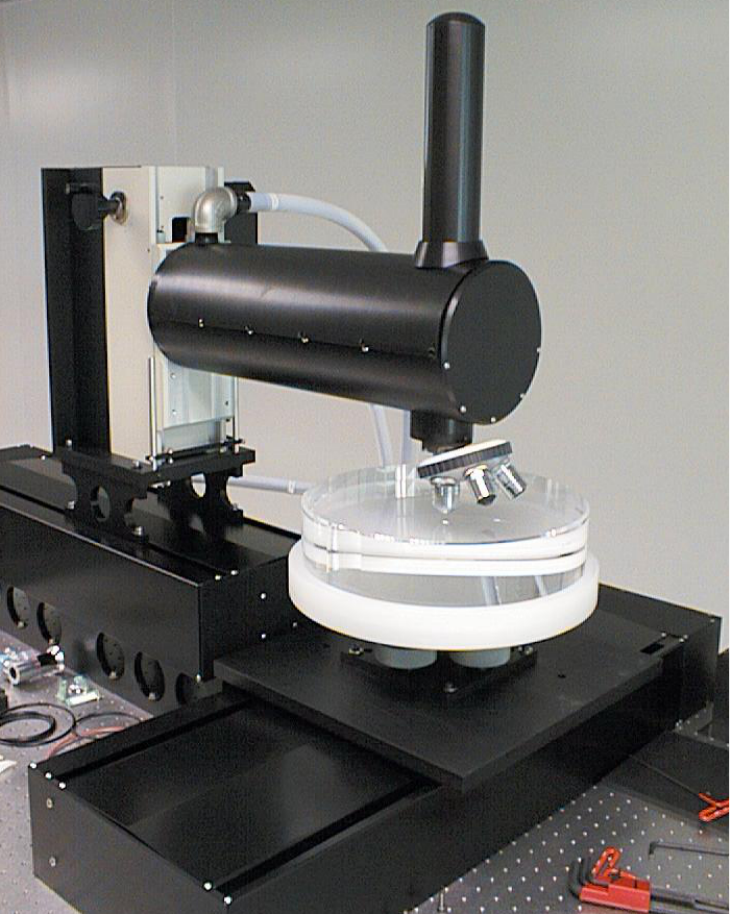}
	\caption{\label{FIGmicromap} (Color online) Micromap optical profiler used for the measurement of the RoC of coated cantilever blades.}
\end{figure}

We used a custom optical profiler developed by Micromap, shown on FIG.~\ref{FIGmicromap}, to measure the RoC of the annealed coated cantilever blades. The instrument is a Mirau interferometer designed specifically for the characterization of large optics, and is equipped with remotely driven micrometric translation stages on the three axes and with manually tunable micrometric screws for tilting (roll and pitch) the sample support plate. With standard settings (monochromatic light at $\lambda = 550$ nm, phase-shifting interferometry), the surface of the sample probed by the laser beam is typically less than 1 mm$^2$.
\begin{figure}
	\includegraphics[width=0.75\textwidth]{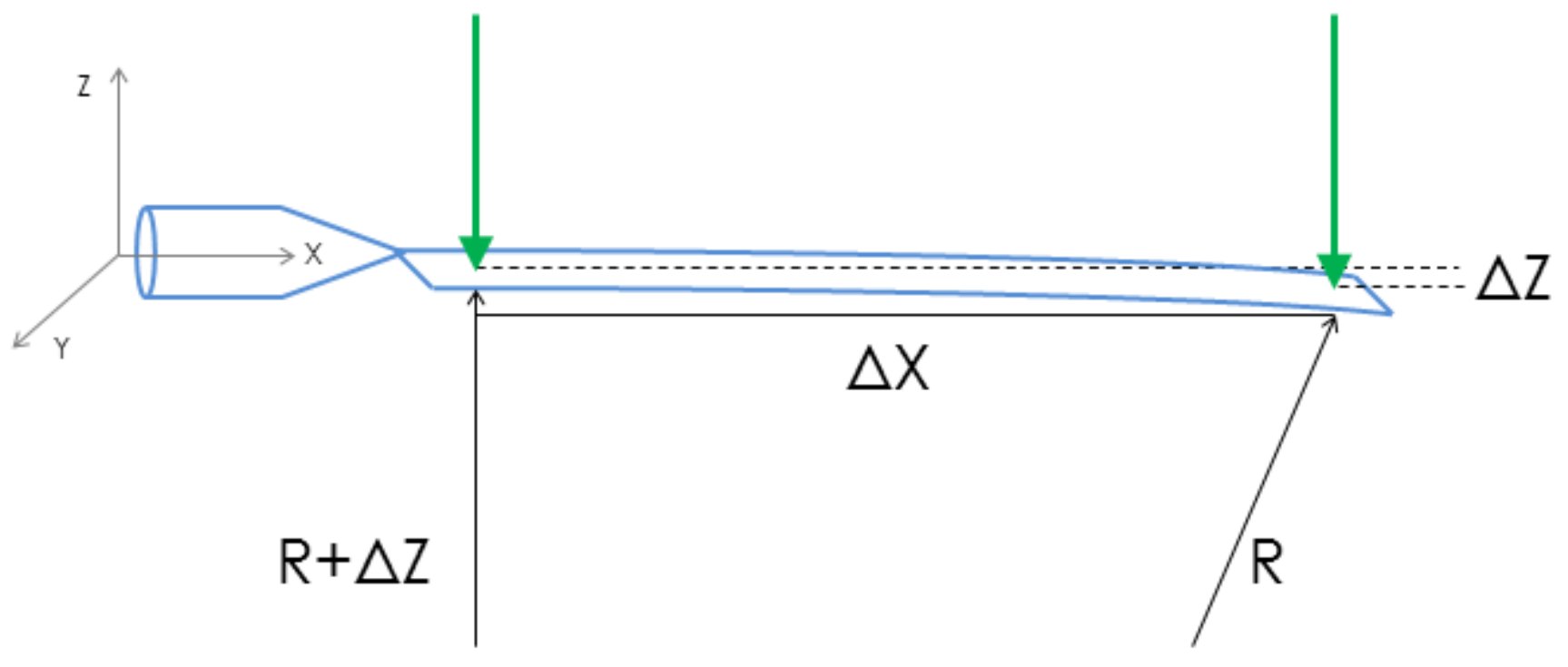}
	\caption{\label{FIGroc} (Color online) Scheme of the technique used to measure the RoC of coated cantilever blades. See Section \ref{SECroc} for more details.}
\end{figure}

The measurement technique is shown on FIG.~\ref{FIGroc}. The longitudinal axis of the cantilever blade (roughly aligned along the abscissa of the Micromap translation stage frame) is swept with steps of a few mm, and at each position the distance to the sample (the $z$ coordinate) is adjusted with steps of 5 $\mu$m to find the focal plane. Once the displacements $\Delta x$ and $\Delta z$ are known, the RoC $R$ of the sample can be readily computed:
\begin{equation}
\label{EQroc}
R = \frac{(\Delta x)^2+(\Delta z)^2}{2\ \Delta x}\ .
\end{equation}
As an estimation of the sample RoC is obtained for each step along $x$, an average RoC with its standard deviation can be calculated.

\subsubsection{Mechanical loss}
\begin{figure}
 \subfigure[]{
	\includegraphics[width=0.48\textwidth]{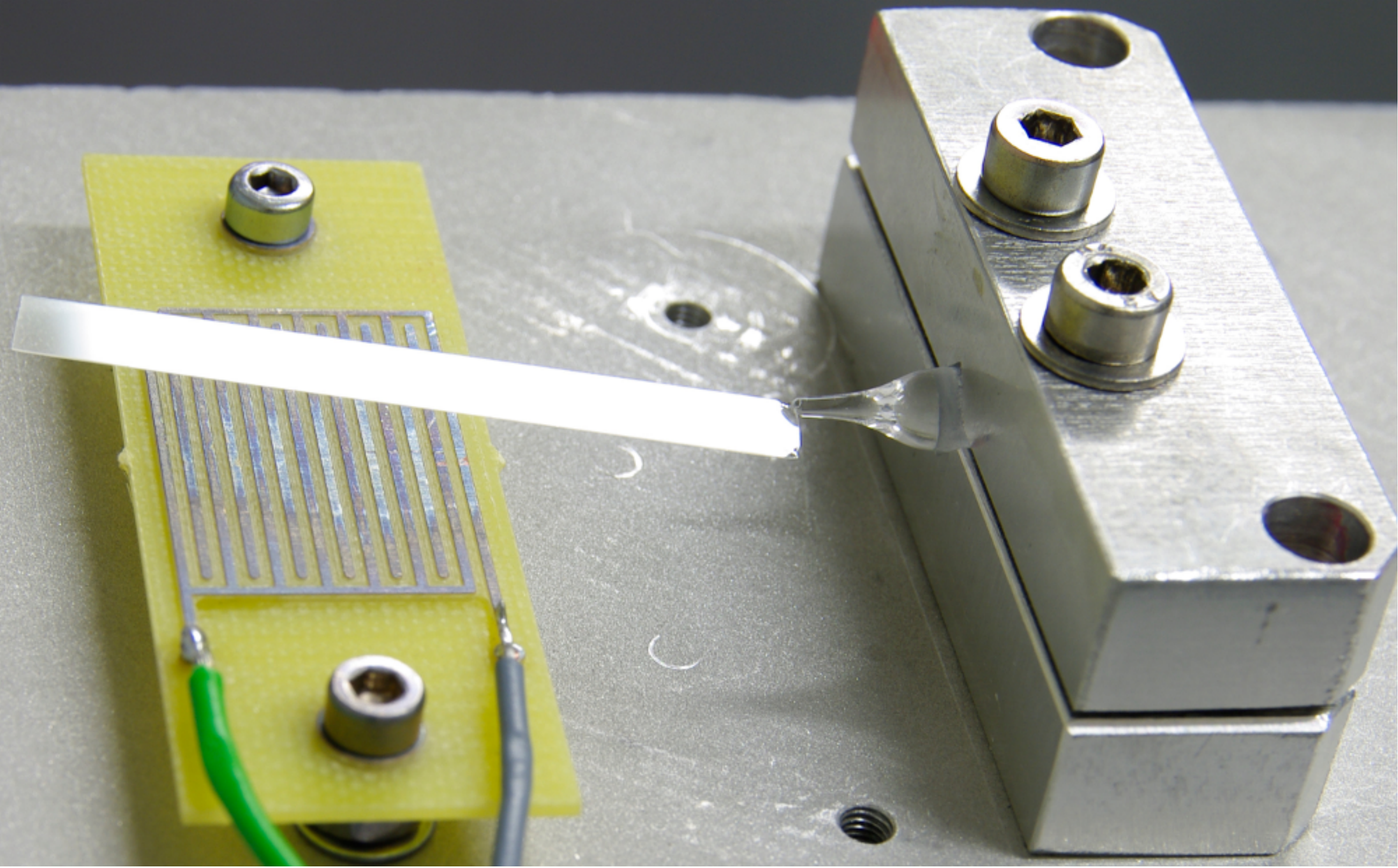}
 }
 \subfigure[]{
   \includegraphics[width=0.48\textwidth]{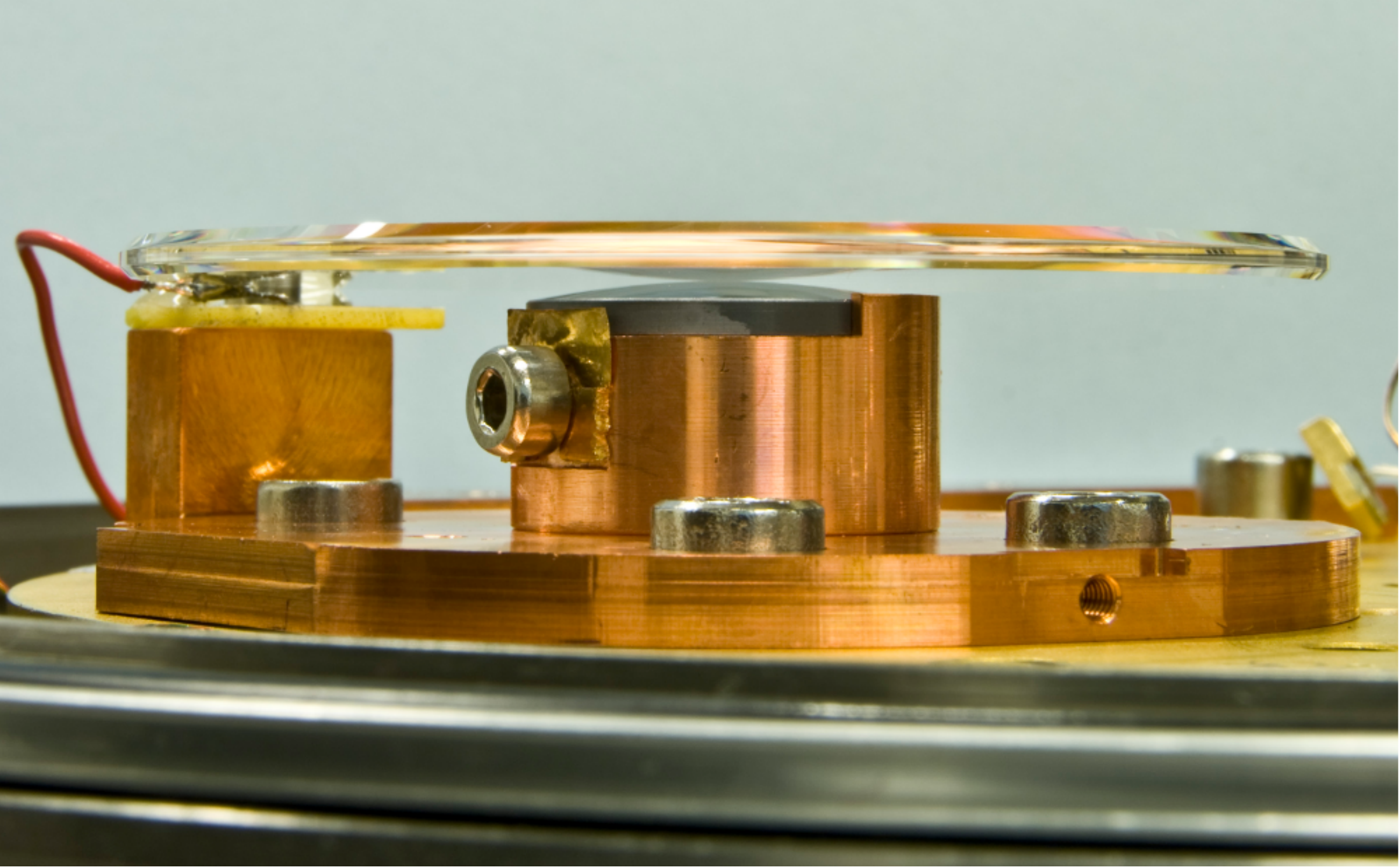}
 }
	\caption{(Color online) Suspension systems for the mechanical loss characterization: (a) clamping of a fused-silica cantilever blade; (b) spherical support of GeNS with a fused-silica disk on top.}
	\label{FIGsuspensions}
\end{figure}

We used two distinct setups, both based on the resonance method, to perform ring-down measurements of all the cantilever blades and disks. The two setups only differ in the system adopted to suspend the samples. In the first apparatus, a clamping system was used to suspend the cantilever blades from the base block. In the second one, a system named Gentle Nodal Suspension \cite{Cesarini09} (GeNS) was used to suspend the disks from the center. Both systems are shown in FIG.~\ref{FIGsuspensions}. In the GeNS system, a sphere provides a mechanically stable support for the disc to be measured, thus minimizing its contact surface. Therefore, the system allows to avoid the clamping of samples, suppresses any potential additional source of damping and exhibits a high reproducibility of the results \cite{Granata15}. All the measurements have been performed in vacuum enclosures at $p \leq 10^{-6}$ mbar, to prevent residual-gas damping.

In both setups, the vibrational modes $\nu_k$ of the resonator are excited with an AC voltage (ranging from 0.2 to 3 kVpp, depending on the sample) applied through a comb-shaped capacitor \cite{Cadez88} placed a few millimeters away from the surface of the sample, in order to avoid any contact. In turn, each mode $k$ is excited well above its ambient noise level, then the excitation is turned off, leaving the resonator free to ring down. The amplitude of the resonating mode is continuously read out through an optical-lever system, where a He-Ne laser is reflected at the surface of the sample towards a quadrant photodiode used as a displacement sensor. A custom-developed software based on LabVIEW is then used to:
\begin{description}
	\item[(i)] acquire the signal of the photodiode
	\item[(ii)] perform its fast Fourier transform and filter it with a narrow (1 Hz) band-pass filter centered around the mode frequency
	\item[(iii)] compute the exponential fit of the envelope $A(t)=A_0\ \text{exp}(-t/\tau_k)$ of the free decay amplitude of the filtered signal
	\item[(iv)] calculate the ring-down time $\tau_k$ and the corresponding loss angle $\phi_k \equiv (\pi\nu_k\tau_k)^{-1}$. 
\end{description}
Typically, we measured three to five modes per sample on cantilever blades, in the 10$^1$-10$^3$ Hz band, and five modes on disks, sampling the 10$^2$-10$^4$ band. As the coating loss is expected to be frequency independent \cite{Saulson90}, the loss angles $\phi_k$ of each sample are averaged to yield a measured loss $\langle \phi_{m} \rangle$ with its standard deviation.

\subsubsection{Morphology}
\begin{figure}
	\includegraphics[width=1.00\textwidth]{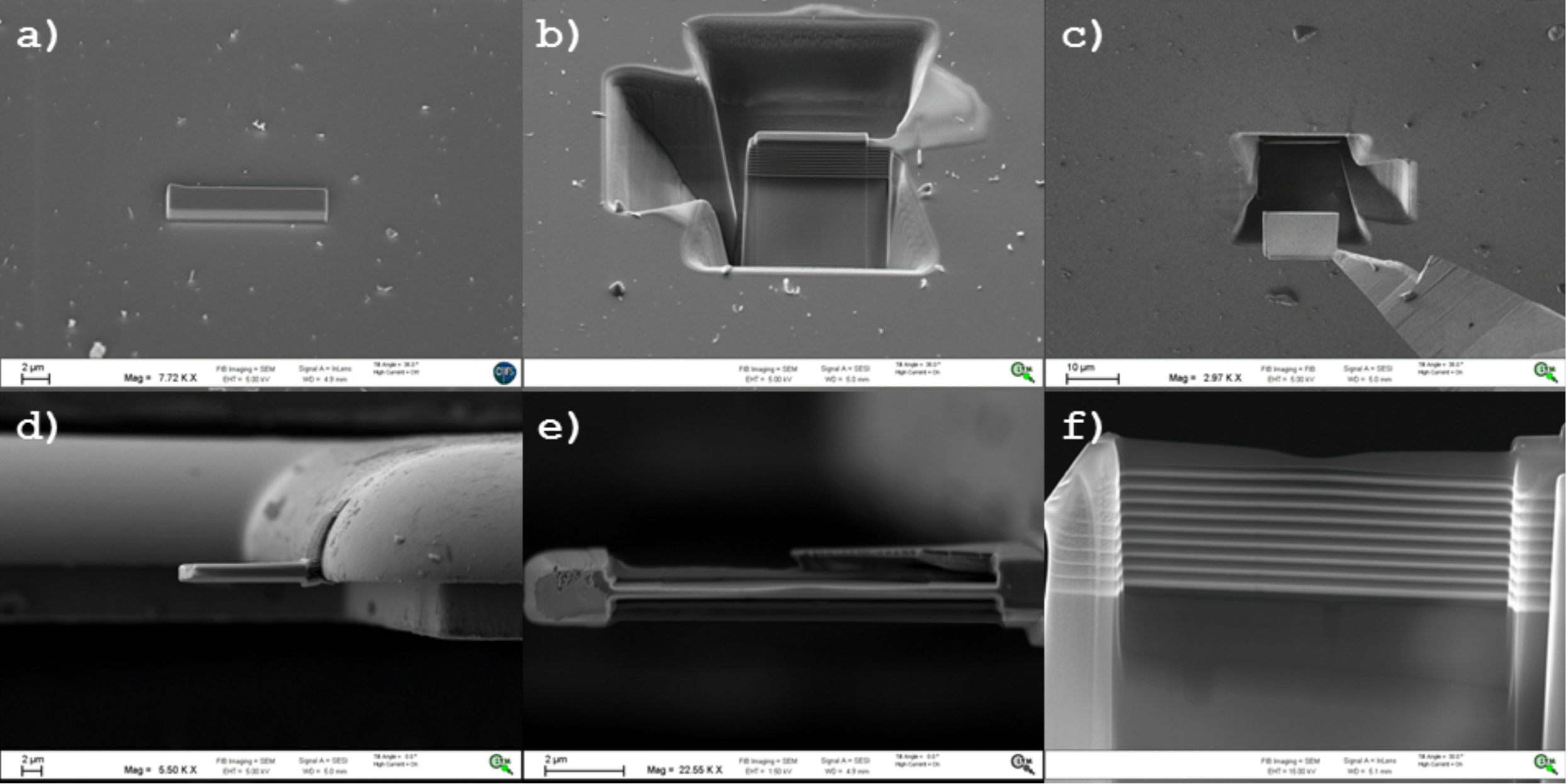}
	\caption{\label{FIGclymFIB} SEM images of the preparation of a thin transverse section of a coating stack through FIB etching: a) chemical-vapor deposition of a 1 $\mu$m-thick protective layer of carbon over the sample to be extracted; b) etching of the coating section by means of a high-energy beam of gallium ions; c) extraction of the section with a tungsten micro-manipulator (tungsten coating for tip welding); d) welding of the section to a copper grating via a carbon coating; e) thinning of the section by means of the high-energy ion beam; f) front view of the thin coating section (bright: TiO$_2$Ta$_2$O$_5$ layers). Courtesy of the CLYM.}
\end{figure}

A JEOL 2010F transmission electron microscope of the Centre Lyonnais de Microscopie\footnote{\url{clym.fr}} (CLYM) has been operated at 200 kV to analyze the structure of a coating stack, and in particular the interface between the layers. To this purpose, a thin transverse section of a stack has been extracted from a test sample using a focused-ion-beam (FIB) setup (ZEISS Nvision 40). The preparation of the sample is depicted in FIG.~\ref{FIGclymFIB}. This sample has then been used for scanning and transmission electron microscopy (SEM and TEM, respectively), energy dispersive x-ray spectrometry (EDX) and electron energy loss spectroscopy (EELS).

\subsection{Simulations}
The calculation of the coating loss requires the knowledge of the dilution factor,
\begin{equation}
\label{EQdilFactor}
D = \frac{E_{c}+E_{s}}{E_{c}}\ ,
\end{equation}
with $E_c$ and $E_s$ being the elastic energy stored in the coating layers and in the substrate, respectively. Preferably, we numerically computed these energies with the finite-element software ANSYS\footnote{\url{ansys.com}}. Alternatively, a simple analytic model is also available \cite{Comtet07}. A comparison between the finite-element simulations and the analytic model is presented in Section \ref{SECexpRes}.

\begin{figure}
 \subfigure[]{
	\includegraphics[width=0.48\textwidth]{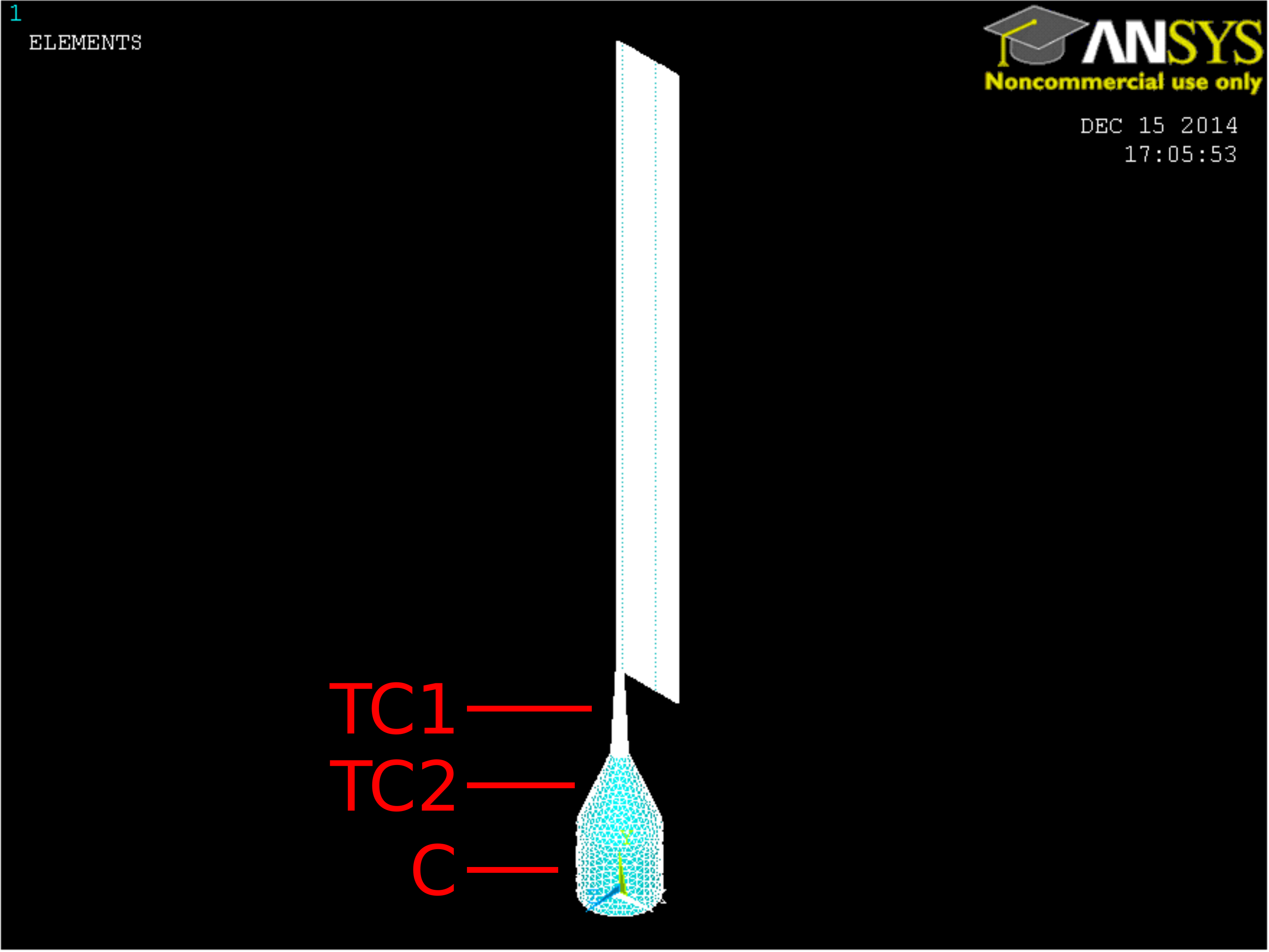}
   	\label{FIGsimBlade}
 }
 \subfigure[]{
   \includegraphics[width=0.48\textwidth]{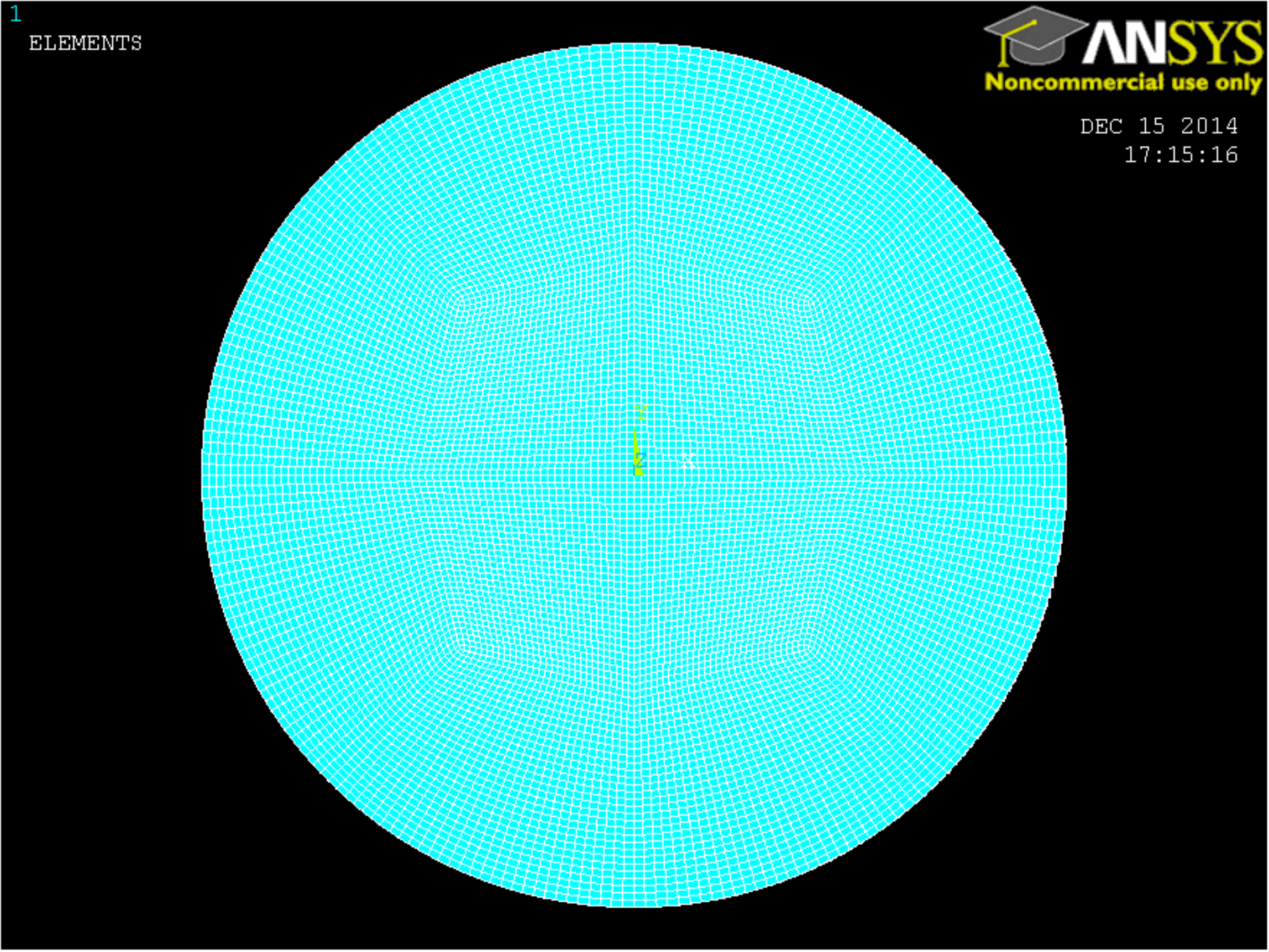}
   \label{FIGsimDisk}
 }
	\caption{(Color online) Meshed models used in finite-element simulations: (a) cantilever blade welded to its base clamping block; (b) disk.}
	\label{FIGsimModels}
\end{figure}

Table \ref{TABsimPar} lists the parameters used in our simulations. To model the coated samples, we used three-dimensional structural elements (\textit{SOLID186}) for the substrates and two-dimensional structural elements (\textit{SHELL281}) for the coatings. Exemplary meshed models are shown on FIG.~\ref{FIGsimModels}.
\begin{table}
	\caption{\label{TABsimPar} Parameters used in the finite-element simulations for bulk and coating materials\footnote{The density of a doped-tantala mono-layer has been measured by means of Rutherford backscattering spectrometry. The density of a silica mono-layer has been deduced from mass measurements and from spectroscopic ellipsometer characterizations of a coated silicon wafer of known dimensions. The refractive index of the coating materials has been estimated from spectrophotometric measurements of transmissivity. The remaining coating parameters have been either taken from the literature (where explicitly mentioned) or, in the case of lack of available information, supposed equal to the homologue parameters of the corresponding bulk materials.}: density $\rho$, Young's modulus $Y$, Poisson ratio $\nu$, shear modulus $G$ and refractive index $n$ at $\lambda_0 = 1064$ nm.}
	\begin{ruledtabular}
	\begin{tabular}{c c c c c c}
		Material & $\rho$ [kg/m$^3$] & $Y$ [GPa] & $\nu$ & $G$ [GPa] & $n$ \\ \hline
		bulk SiO$_2$ 				& 2200	& 73.2						& 0.164 			& 				&  \\
		bulk Si	\cite{Hopcroft10}	& 2330 	& $Y_x = 169.7$				& $\nu_{xy}=0.061$	& $G_{xy}=51$ 	&  \\
									&									& $Y_y = 169.7$	& $\nu_{yz}=0.278$	& $G_{yz}=80$ 	&  \\
									&									& $Y_z = 130.4$	& $\nu_{xz}=0.362$	& $G_{xy}=80$ 	&  \\
		coating TiO$_2$Ta$_2$O$_5$	& 6505	& 140 \cite{Abernathy14}	& 0.230				&				& 2.070 \\
		coating SiO$_2$				& 2470	& 73.2						& 0.164				&				& 1.442 \\		
	\end{tabular}
	\end{ruledtabular}
\end{table}

\begin{figure}
 \subfigure[]{
	\includegraphics[width=0.48\textwidth]{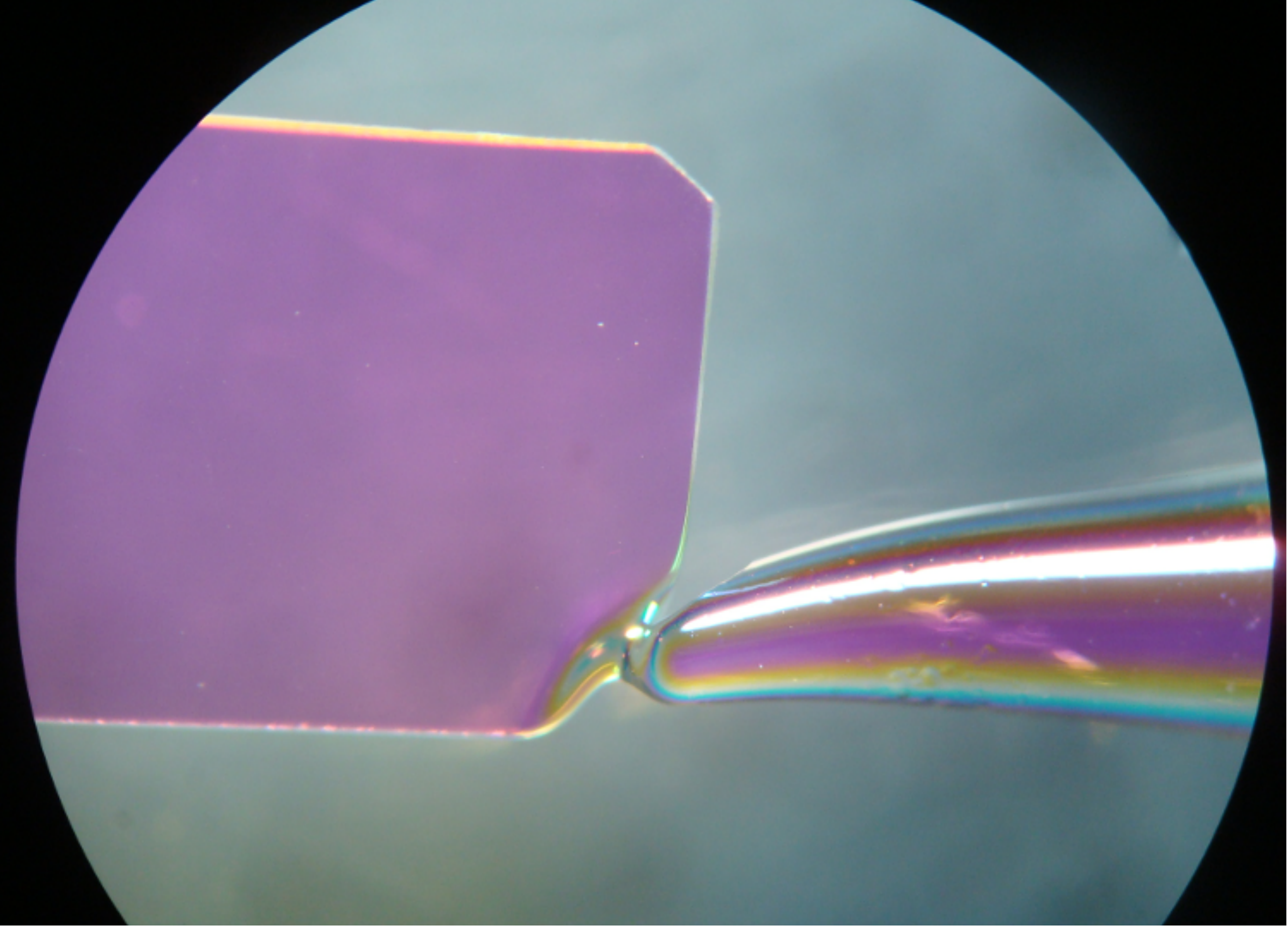}
 }
 \subfigure[]{
   \includegraphics[width=0.46\textwidth]{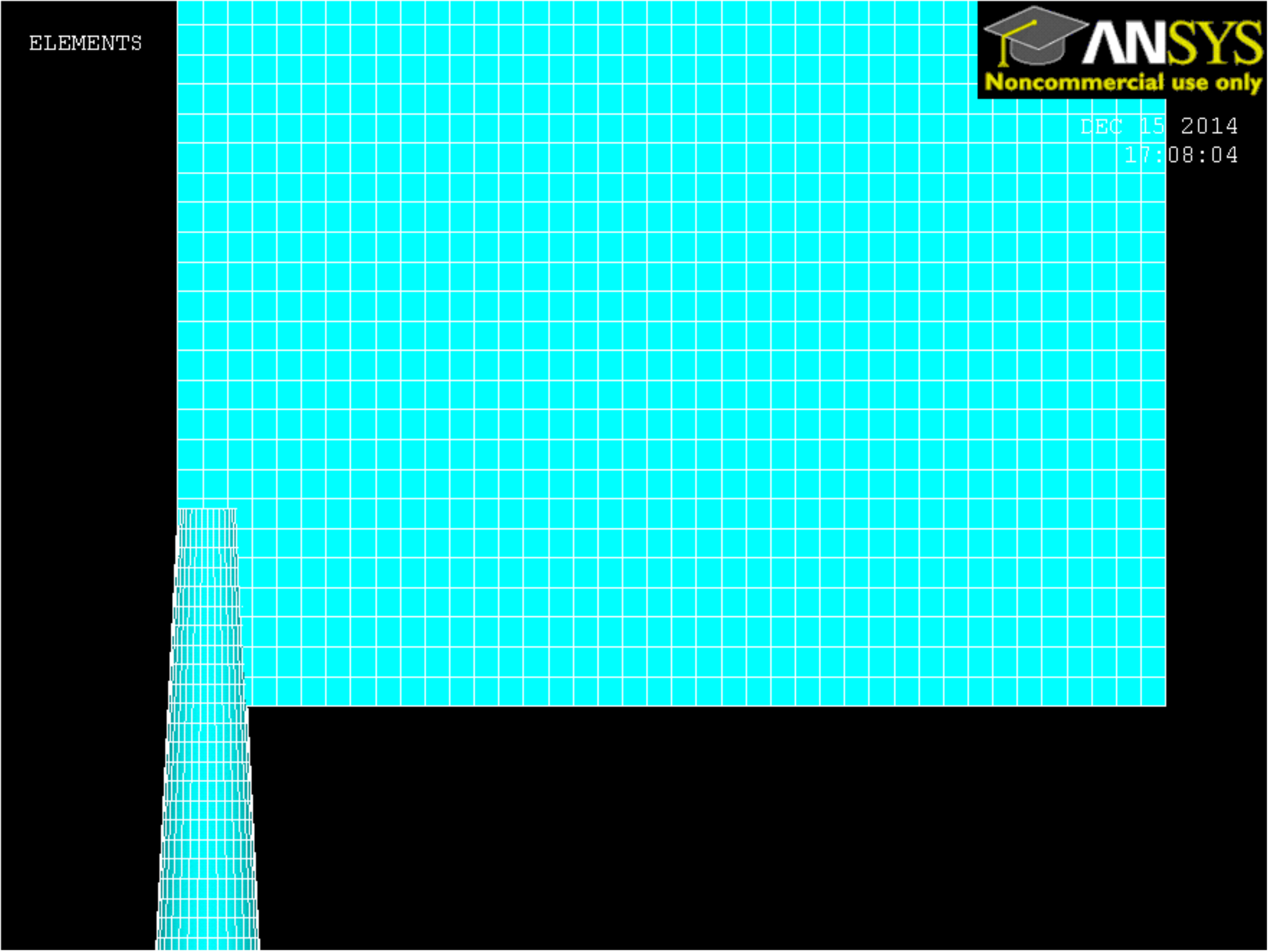}
 }
	\caption{(Color online) Different welding points at the base of fused-silica cantilevers: (a) photograph of a coated sample with point-like contact; (b) finite-element meshed model with maximum overlap ($l=1$ mm).}
	\label{FIGwelding}
\end{figure}

\begin{figure}
	\includegraphics[width=0.25\textwidth]{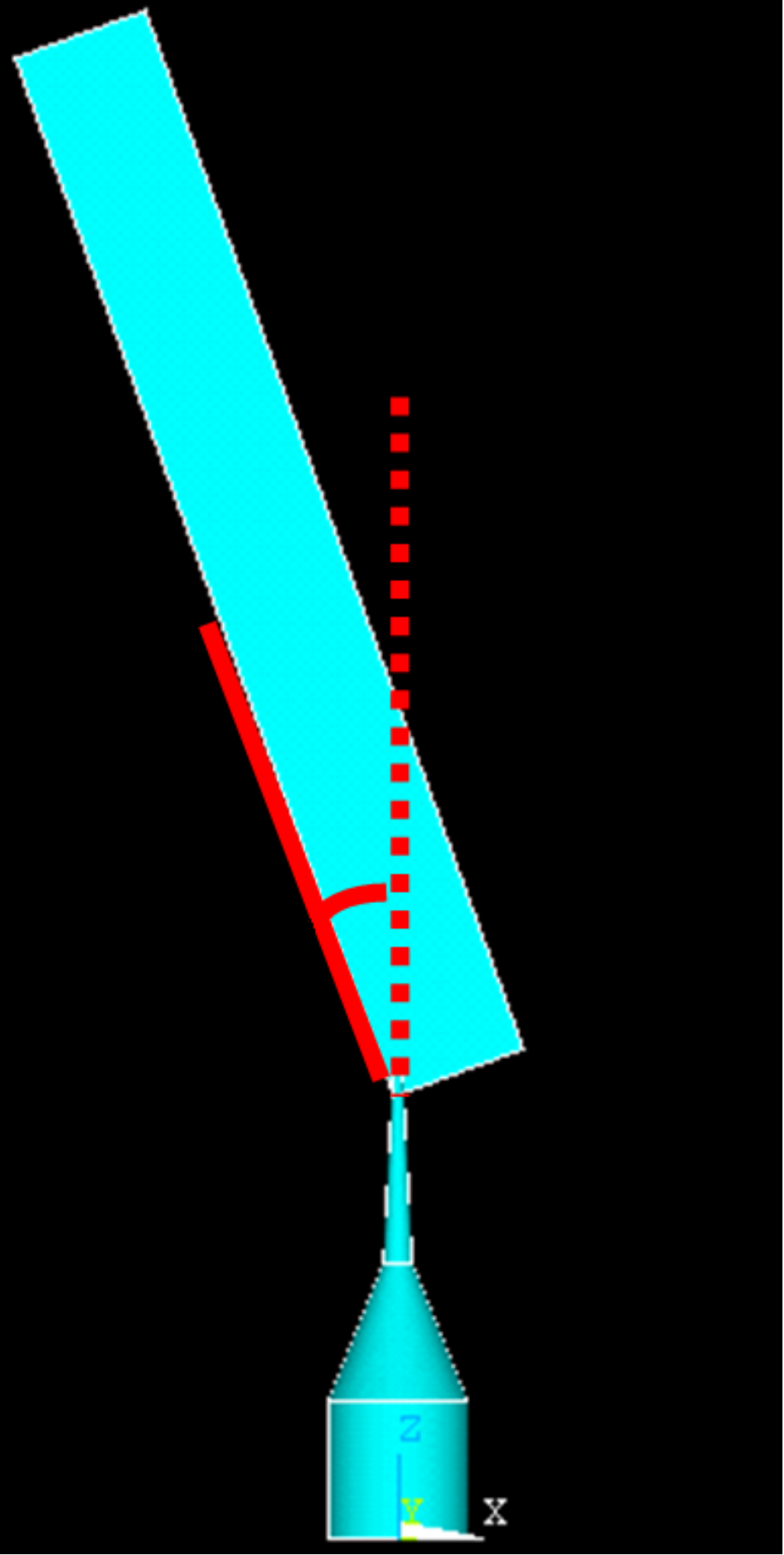}
	\caption{\label{FIGweldAngle} (Color online) Simulation of a fused-silica cantilever welded to an angle $\theta$ to its clamping block.}
\end{figure}

Special care was taken to accurately simulate the geometry of the welding point of cantilever blades, modeled as a volume of a truncated cone (TC1) which is partially overlapping with the volume of the cantilever blade itself. TC1 is part of the clamping block model, composed of a second adjacent truncated cone (TC2) and a short terminal cylinder (C). The diameter at the base of TC1 is the same at the top of TC2, and the same holds for TC2 and C. The simulated weld joint is shown on FIG.~\ref{FIGwelding}, where it is compared to a photograph of a welded cantilever blade. Several configurations were tested, to choose the model yielding the closest results to the measured resonant frequencies. We independently changed the following:  
\begin{itemize}
	\item the overlap length $l$ between TC1 and the cantilever blade, set either to 0 (point-like contact) or to 1 mm
	\item the angle $\theta$ between the longitudinal axis of the cantilever blade and the rotation axis of the clamping block, in the plane of the blade, as shown on FIG.~\ref{FIGweldAngle}
	\item the size of TC1, by changing its base and top diameters from $d_{1b} = 1$ to $d_{1b} = 1.5$ mm and from $d_{1t} = 0.3$ to $d_{1t} = 1.2$ mm, respectively.
\end{itemize}
As meshing is a critical step of the simulations, we set a fine mesh for the cantilever blade (typical mesh size values are 150 $\mu$m for length and width, 50 $\mu$m for thickness) and for its clamping block (100 $\mu$m for TC1 and 500 $\mu$m for TC2 and C), shown on FIG.~\ref{FIGsimBlade}. The welding was simulated by requiring that adjacent nodes of overlapping and contiguous volumes be merged together.

The welding angle $\theta$ and the diameters ($d_{1b}$ and $d_{1t}$) of TC1 proved to have a small influence on the value of the simulated resonant frequencies, yielding a variation of the order of $<2$\% and $<5$\%, respectively. The overlap length $l$ came out instead to be a critical parameter, changing the mode frequencies as much as 40\% (depending on the mode). The best agreement between the measured and the simulated resonances was obtained with $l = 1$ mm, $d_{1b} = 1$ mm and $d_{1t} = 0.3$ mm. For simplicity, we also imposed $\theta = 0$ in the same parameter set. This model provides results which are consistent with the measured resonant frequencies of a bare substrate within a 0.1-2\% error, depending on the mode. By also including the coating in the simulations, the model is in agreement with the measured resonant frequencies of a coated substrate within a 2-4\% error. 
Concerning the dilution factor, the results showed that its value does not depend on any of the above-mentioned model parameters. This may be explained by the fact that, according to our simulations, the typical distribution of the elastic energy in the vibrating coated cantilever blade is the following: less than $0.5$\% is in the whole clamping block (TC1 + TC2 + C)\footnote{In this configuration TC2 has base and top diameters equal to $d_{2b} = 1.5$ mm and $d_{2t} = 1.2$ mm, respectively, C has radius $r = 2.5$ mm and height $h = 5$ mm.}, then $\sim$85-95\% is in the substrate and $\sim$5-15\% is in the coating stack (depending on the coating thickness). This result has also been confirmed by a further test, where we simulated only a cantilever blade clamped along its base. Also in this case, the dilution factor turned out to be the same. Furthermore, we found out that the dilution factor is mode independent, no matter the model used to simulate the cantilever blades.

In the case of mono-layers, the {\it specific} dilution factors (i.e. independent from coating thickness) are 37520 nm for silica coatings and 18830 nm for titania-doped tantala coatings.  For a given substrate configuration, specific dilution factors are more useful as they can be used to calculate the dilution factors of mono-layers of any desirable thickness.

For disks, we set a mesh size of 1 mm for in-plane surface and volume elements. Along the direction of the thickness $t$, we set a mesh size roughly equal to $t/2$. For these samples the dilution factor actually depends on the mode shape, which can be classified with respect to the number of radial and azimuthal nodes. We denote by $(r, a)_k$ the $k$-th mode with $r$ radial and $a$ azimuthal nodes. In Table \ref{TABspecDilFact} are the specific dilution factors for silica and titania-doped tantala mono-layers on a 2.54 mm-thick disk of SiO$_2$, as a function of the mode order.
\begingroup
\squeezetable
\begin{table}
	\caption{\label{TABspecDilFact} Mode-dependent specific dilution factors $D_k$ for mono-layers deposited on a $\varnothing\,$ 3", 2.54 mm-thick disk of fused silica, from finite-element simulations: frequency $f$, mode order\footnote{$(r, a)_k$ is the $k$-th mode with $r$ radial and $a$ azimuthal nodes.} $(r, a)_k$, factors for silica and titania-doped tantala coatings.}
	\begin{ruledtabular}
	\begin{tabular}{c l c c}
		$f$ [kHz] & $(r, a)_k$ & $D_k$ - SiO$_2$ [nm] & $D_k$ - TiO$_2$Ta$_2$O$_5$ [nm] \\ \hline
		2.7 & $(0, 2)_1$ & 213981 & 116099 \\
		6.2 & $(0, 3)_4$ & 215272 & 115607 \\
		10.7 & $(0, 4)_8$ & 216748 & 115668 \\
		16.1 & $(0, 5)_{10}$ & 218408 & 116088 \\
		22.5 & $(0, 6)_{15}$ & 220437 & 116773 \\
		24.2 & $(1, 3)_{17}$ & 220806 & 113004 \\
		29.8 & $(0, 7)_{21}$ & 222835 & 117662 \\
		33.2 & $(1, 4)_{23}$ & 223573 & 114763 \\
		37.7 & $(2, 2)_{27}$ & 224126 & 113898 \\
		37.8 & $(0, 8)_{29}$ & 225233 & 118717 \\
		43.1 & $(1, 5)_{34}$ & 226709 & 116491 \\		
		46.5 & $(0, 9)_{36}$ & 227816 & 119906 \\
		49.2 & $(2, 3)_{39}$ & 227447 & 115923 \\
	\end{tabular}
	\end{ruledtabular}
\end{table}
\endgroup
\subsection{Model}
\label{SEClossMod}
For a coated cantilever blade, the equation of the total loss in its simplest form reads \cite{Crooks02, Harry02}
\begin{equation}
\label{EQsimpleMod}
\phi_k^{cs} = \frac{1}{D} \left[ (D-1)\phi_k^s + \phi_k^c \right] \ ,
\end{equation}
where $D$ is the dilution factor defined in Eq.(\ref{EQdilFactor}), $\phi^{cs}_k$ is the measured loss of the coated sample, $\phi^s_k$ is the measured loss of the bare substrate and $\phi_k^c$ is the coating loss. In this model, any additional systematic loss (such as clamp-surface friction, residual-gas damping, etc.) is assumed to be negligible.

We could safely use Eq.(\ref{EQsimpleMod}) in our analysis, for the presence of a clamping block at the base of the cantilever blades largely reduces the clamp loss, and the GeNS system yields negligible excess damping by design. However, we wanted to modify Eq.(\ref{EQsimpleMod}) in order to test the presence in our data of a loss at the interface between the stack layers (or between the coating and the substrate in mono-layers). 

We actually did not develop the model of the interface loss. One could suppose that the loss depends on the strain gradient along the direction normal to the coating, or that the loss is related to a change in the structure of the interface layer, whose thickness measurement is reported in Section \ref{SECexpRes}. In order to overcome these difficulties, we decided to include an additive interface loss $\phi_i$ in the total loss of the coated oscillator, where the unknown dilution factor for this additional loss is included inside $\phi_i$ itself. We also assume that the energy $E_i$ of the interfaces is negligible compared to the elastic energy of the substrate and of the coating bulk. Finally, $\phi_i$ is assumed to be mode independent\footnote{This assumption seems reasonable for cantilever blades, whereas for disks this might not be true.}. With these hypotheses, our modified model writes
\begin{equation}
\label{EQinterfMod}
\phi_k^{cs} = \frac{1}{D} \left[ (D-1)\phi_k^s + \tilde{\phi}_k^c \right] + N\phi_i \ ,
\end{equation}
with $\tilde{\phi}_k^c$ being the coating bulk loss. By comparison of Eqs.(\ref{EQsimpleMod}) and (\ref{EQinterfMod}), the total coating loss angle is
\begin{equation}
\label{EQphiCmod}
\phi_k^c = \tilde{\phi}_k^c + N D \phi_i \ ,
\end{equation}
and so the outcome of the measurements, the frequency-averaged measured coating loss angle, is
\begin{equation}
\label{EQmeasPhiCmod}
\langle \phi_m \rangle \equiv \frac{1}{m} \sum_{k=1}^m \left[ D\phi^{cs}_k - (D-1)\phi^{s}_k \right] = \frac{1}{m} \sum_{k=1}^m \tilde{\phi}^c_k + ND \phi_i\ .
\end{equation}
For disks, Eq.(\ref{EQmeasPhiCmod}) is slightly different in that $D \equiv D_k$.

We will see in the following that our data might include an interface loss, though unfortunately our set of samples is not optimal to draw any definitive conclusion on this subject. Instead, previous works \cite{Penn03} have demonstrated the absence of an interface loss through the measurement of different coatings.

As it will play a crucial role in our analysis, we recall here the equation of the expected coating loss \cite{Penn03} of a stack, \begin{equation}
\label{EQphiE}
\phi_e = \frac{\sum_{j} t_j Y_j \langle \phi_m \rangle_j}{\sum_{j} t_j Y_j}\ ,
\end{equation}
which is defined as the linear combination of the loss measured on mono-layers: given the $j$-th material of the stack ($j= \text{SiO$_2$}, \text{TiO$_2$Ta$_2$O$_5$}$), $t_j$ is its total thickness, $Y_j$ its Young's modulus and $\langle \phi_m \rangle_j$ its mechanical loss. The thickness ratio $r = t_{\text{TiO$_2$Ta$_2$O$_5$}}/t_{\text{SiO$_2$}}$ is also an important parameter to qualify each particular stack with respect to its expected loss.

\subsection{Results}
\label{SECexpRes}

\subsubsection{Mono-layers}
\begin{figure}
	\centerline{\includegraphics[width=0.75\textwidth]{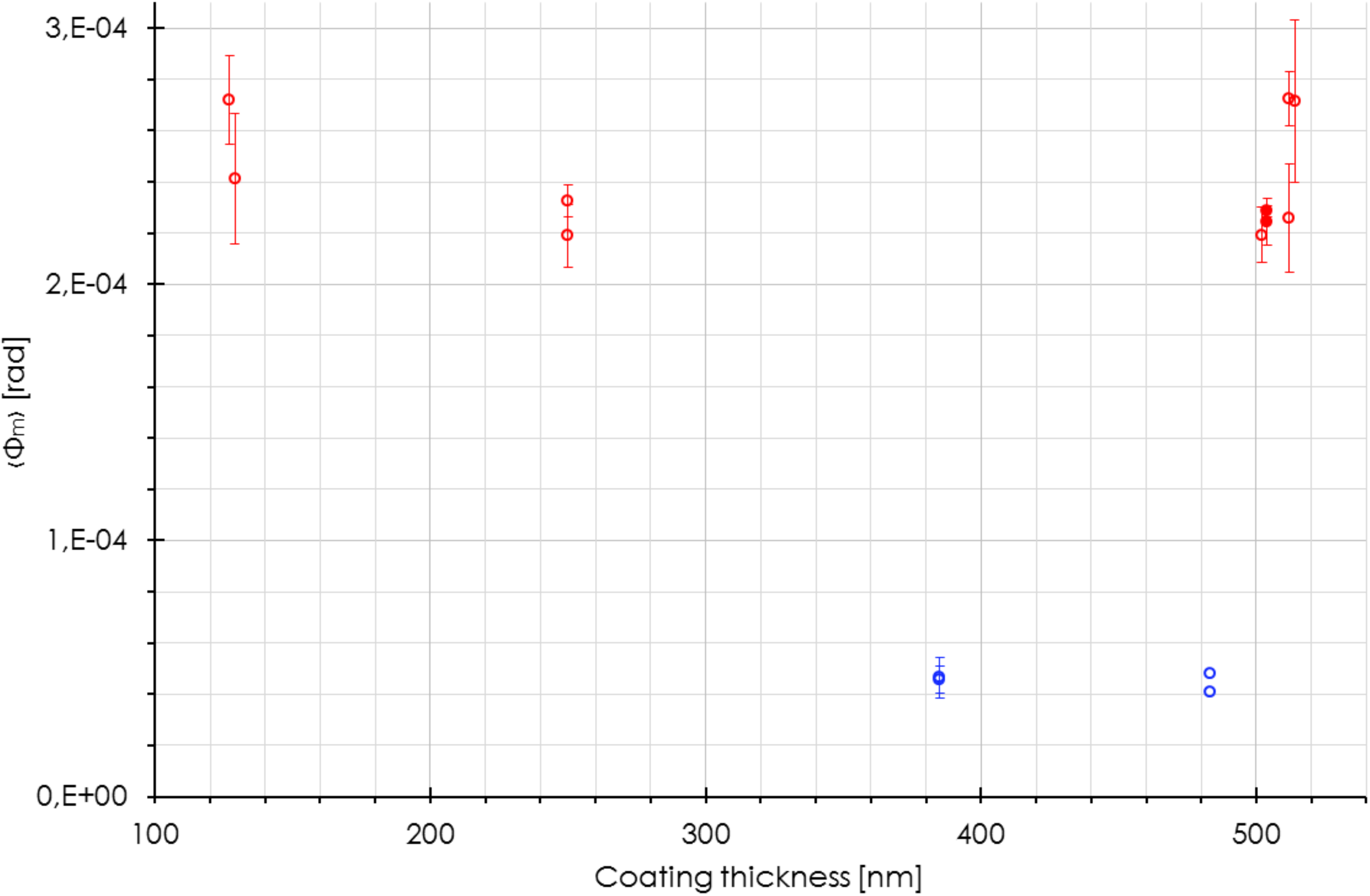}}
	\caption{\label{FIGlossMonol} (Color online) Mechanical loss of mono-layers of silica (blue) and titania-doped tantala (red) as a function of the layer thickness, as measured on fused-silica cantilever blades coated on one (open circles) or on both surfaces (dots).}
\end{figure}

The mechanical loss of mono-layers deposited on fused-silica cantilever blades can be computed via Eq.(\ref{EQmeasPhiCmod}). Since in this case $N=1$, we expect $\langle \phi_{m} \rangle$ to be dependent of the coating thickness via the dilution factor $D$. The results are shown on FIG.~\ref{FIGlossMonol} and listed in Table \ref{TABmonolayers}.
\begingroup
\squeezetable
\begin{table}
	\caption{\label{TABmonolayers} Specifications and results of fused-silica cantilever blades coated with mono-layers: sample, coating material, coating  thickness $t$, simulated and analytic dilution factors $D$ and $D_{an}$, respectively, measured coating loss $\langle \phi_{m} \rangle$ in units of $10^{-4}$ rad.}
	\begin{ruledtabular}
	\begin{tabular}{l c c c c c}
		sample\footnotemark[1] & coating & $t$ [nm] & $D$ & $D_{an}$ & $\langle \phi_{m} \rangle$ \\ \hline
		w8		& TiO$_2$Ta$_2$O$_5$ & 513 & 		& 37.7 	& 2.4 $\pm$ 0.2 \\
		151 	& TiO$_2$Ta$_2$O$_5$ & 514 & 		& 37.7 	& 2.7 $\pm$ 0.3 \\
		190 	& TiO$_2$Ta$_2$O$_5$ & 502 & 		& 38.6	& 2.2 $\pm$ 0.1 \\
		GH-A1	& TiO$_2$Ta$_2$O$_5$ & 512 & 		& 37.8 	& 2.7 $\pm$ 0.1 \\
		GH-A4	& TiO$_2$Ta$_2$O$_5$ & 512 & 		& 37.8	& 2.3 $\pm$ 0.2 \\
		GH-A5	& TiO$_2$Ta$_2$O$_5$ & 127 & 149.6 	& 149.1	& 2.7 $\pm$ 0.2 \\
		GH-A6	& TiO$_2$Ta$_2$O$_5$ & 129 & 		& 147.0	& 2.4 $\pm$ 0.3 \\
		GH-A7	& TiO$_2$Ta$_2$O$_5$ & 250 & 77.7	& 74.8 	& 2.3 $\pm$ 0.1 \\
		GH-A8	& TiO$_2$Ta$_2$O$_5$ & 250 & 		& 74.8 	& 2.2 $\pm$ 0.1 \\
		GH-A7 	& 2 $\times$ TiO$_2$Ta$_2$O$_5$ \footnotemark[2] & 250/254 & & 38.3 & 2.3 $\pm$ 0.1 \\
		GH-A8	& 2 $\times$ TiO$_2$Ta$_2$O$_5$ \footnotemark[2] & 250/254 & & 38.3 & 2.2 $\pm$ 0.1 \\
		H-1		& SiO$_2$ & 483 & 78.0	& 80.4 & 0.48 \\
		H-2		& SiO$_2$ & 483 & 		& 80.4 & 0.41 \\
		GH-A9	& SiO$_2$ & 385 & 97.3	& 96.3 & 0.47 $\pm$ 0.08 \\
		GH-A10	& SiO$_2$ & 385 & 		& 96.3 & 0.46 $\pm$ 0.05 \\
	\end{tabular}
	\end{ruledtabular}
	\footnotetext[1]{samples with the same coating specifications have been coated together, by pairs.}
	\footnotetext[2]{both surfaces have been coated.}
\end{table}
\endgroup
In the limit of the experimental uncertainties, these data show no significant dependency on the layer thickness. This implies that for mono-layers $\phi_i = 0$, i.e. the loss at the interface between the substrate and the coating is negligible. This is in agreement with the outcome of previous works on different coatings \cite{Penn03}.

A further average of the measured loss over each set of samples yields $\langle \phi_m \rangle_{\text{SiO}_2} = 4.5 \pm 0.3 \times 10^{-5}$ and $\langle \phi_m \rangle_{\text{TiO}_2\text{Ta}_2\text{O}_5} = 2.4 \pm 0.3 \times 10^{-4}$ for the silica and titania-doped tantala coatings, respectively. The silica loss is in agreement with previous estimations from resonant measurements on suspended fused-silica disks \cite{Penn03} and from direct broadband interferometric measurements of thermal noise on coating mono-layers \cite{Li14} and stacks \cite{Principe15}. Regarding the titania-doped tantala loss (already demonstrated in 2010 \cite{Flaminio10}), however, there is a discrepancy between our result and the loss estimated through interferometric measurements of the thermal noise of coating stacks (yielding $3.7 \pm 0.3 \times 10^{-4}$) \cite{Principe15}. The reasons for this discrepancy are unknown to date.

To test the influence of internal stresses on mechanical loss, two cantilever blades (GH-A7 and GH-A8) have also been coated on their second surface with a nominally identical layer, after their first measurement. The second coating is expected to compensate the curvature of the sample due to the first coating, thus determining a new equilibrium of stresses. However, according to our measurements, the second coating has no influence on the measured loss of mono-layers. In the limit of thin mono-layers, either the loss is stress independent or the effect is too small to be detected (given the precision of our measurements).

\subsubsection{Stacks}
\begin{figure}
	\includegraphics[width=0.75\textwidth]{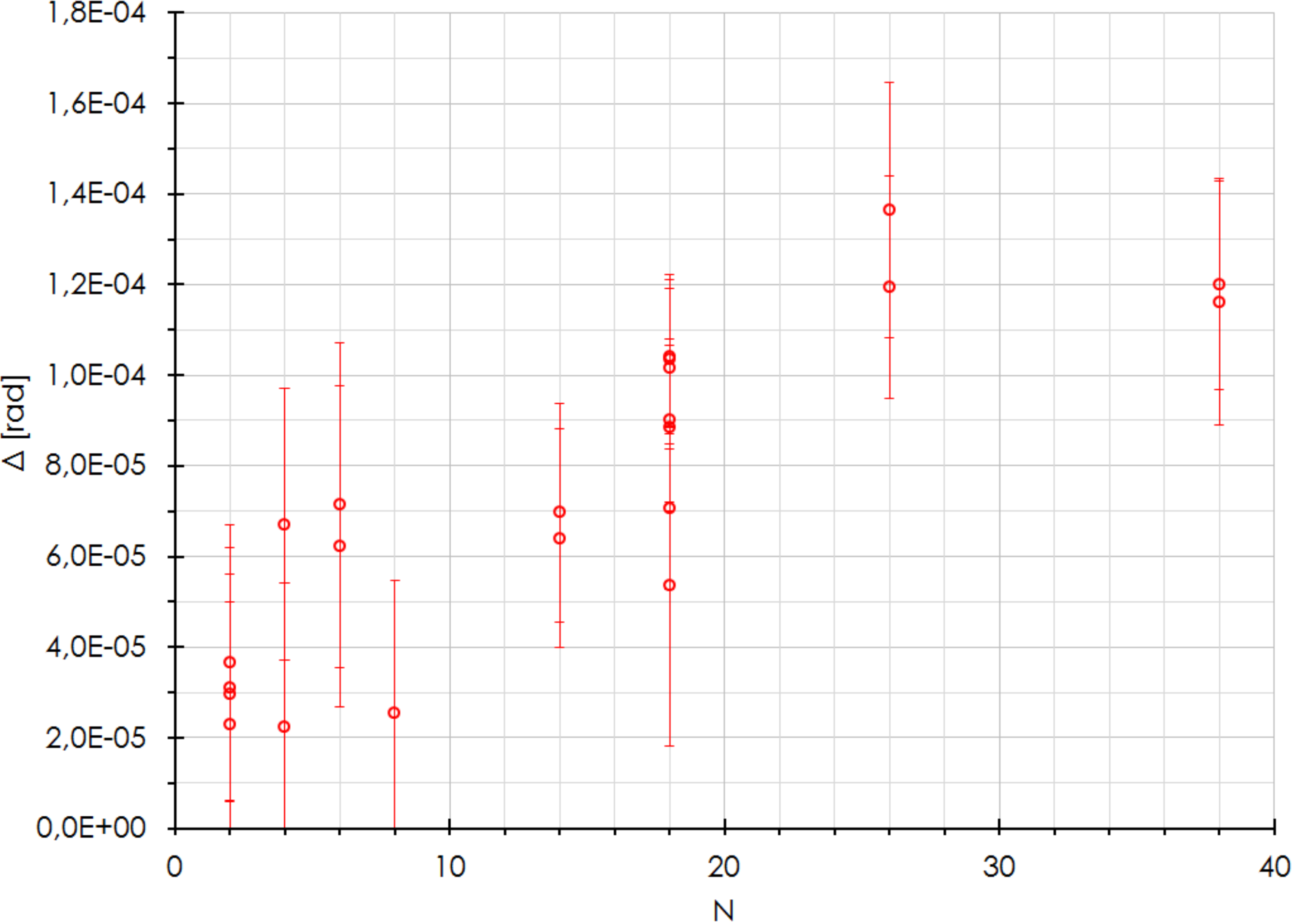}
	\caption{\label{FIGdeltaVSn} Excess loss $\Delta$ of coating stacks deposited on cantilever blades, as a function of the number $N$ of layers in the stack. Excess loss is observed, in violation of Eq.(\ref{EQsimpleMod}).}
\end{figure}

\begingroup
\squeezetable
\begin{table}
	\caption{\label{TABstacksBlades} Specifications and results of fused-silica cantilever blades coated with stacks: sample, coating design (HR = high reflection with optimized thickness, AR = anti-reflection, QWL = quarter wavelength), number of coating layers $N$, laser wavelength $\lambda_0$, thickness ratio $r$, simulated and analytic dilution factors $D$ and $D_{an}$, respectively, measured RoC $R$, expected coating loss $\phi_{e}$, measured coating loss $\langle \phi_{m} \rangle$. Loss angles are given in units of $10^{-4}$ rad.}
	\begin{ruledtabular}
	\begin{tabular}{l c c c c c c c c c}
		sample\footnotemark[1] & coating & $N$ & $\lambda_0$ [nm] & $r$ & $D$ & $D_{an}$ & $R$ [m] & $\phi_{e}$ & $\langle \phi_{m} \rangle$ \\ \hline
		192 		& HR 	& 26 & 1064 & 0.61 & 9.2	& 8.2	& -0.70$\pm$0.05 & 1.5$\pm$0.2 & 2.9$\pm$0.2 \\
		193 		& HR 	& 26 & 1064 & 0.61 & 8.9 	&		& -0.70$\pm$0.06 & 1.5$\pm$0.2 & 2.7$\pm$0.1 \\
		201 		& QWL	& 14 & 1529	& 0.62 & 10.1 	& 9.1	& -1.80$\pm$0.12 & 1.5$\pm$0.2 & 2.2$\pm$0.1 \\
		202 		& QWL	& 14 & 1529	& 0.62 & 9.8 	&		& -1.71$\pm$0.17 & 1.5$\pm$0.2 & 2.2$\pm$0.1 \\
		H-5			& HR	& 18 & 1064 & 0.35 & 12.2 	& 12.6	& 				 & 1.3$\pm$0.2 & 2.2$\pm$0.1 \\
		H-6			& HR	& 18 & 1064 & 0.35 & 12.2 	&		& 				 & 1.3$\pm$0.2 & 2.1$\pm$0.1 \\
		GH-A2		& HR	& 18 & 1064 & 0.32 & 11.7 	& 11.0	& -0.58$\pm$0.01 & 1.2$\pm$0.2 & 2.2$\pm$0.1 \\
		GH-A3		& HR	& 18 & 1064 & 0.32 & 11.7 	&		& 				 & 1.2$\pm$0.2 & 2.3$\pm$0.1 \\
		GH-A11		& QWL	& 2	 & 1064	& 0.70 & 86.7 	& 87.9	& 				 & 1.6$\pm$0.3 & 1.9$\pm$0.1 \\
		GH-A12		& QWL	& 2	 & 1064	& 0.70 & 87.3 	&		& 				 & 1.6$\pm$0.3 & 1.8$\pm$0.1 \\
		GH-A13		& 		& 2  & 1064	& 1.03 & 26.6 	& 26.0	& -2.34$\pm$0.06 & 1.8$\pm$0.3 & 2.1$\pm$0.2 \\
		GH-A15		& 		& 2  & 1064	& 1.03 & 26.6 	&		& -2.28$\pm$0.09 & 1.8$\pm$0.3 & 2.1$\pm$0.1 \\
		HTM-A2		& 		& 8  & 1064	& 1.03 & 26.6 	& 25.9	& -1.73$\pm$0.06 & 1.8$\pm$0.3 & 2.0$\pm$0.1 \\
		HTM-A3		& 		& 18 & 1064	& 1.03 & 12.8 	& 12.1	& -1.41$\pm$0.07 & 1.8$\pm$0.3 & 2.3$\pm$0.2 \\
		HTM-A106	& HR	& 18 & 1064	& 0.33 & 12.2 	& 11.6	& -0.80$\pm$0.05 & 1.2$\pm$0.2 & 1.9$\pm$0.1 \\
		HTM-A107	& HR	& 18 & 1064	& 0.33 & 12.2 	&		& -0.68$\pm$0.03 & 1.2$\pm$0.2 & 2.3$\pm$0.1 \\
		HTM-A119	& AR	& 4	 & 1064 & 1.08 & 25.8 	&		& -1.23$\pm$0.05 & 1.8$\pm$0.3 & 2.5 \\
		HTM-A120	& AR	& 4	 & 1064 & 1.08 & 25.8 	&		& -1.38$\pm$0.08 & 1.8$\pm$0.3 & 2.0$\pm$0.1 \\
		HTM-A121	& AR	& 6	 & 1064 & 1.46 & 16.8 	&		& -1.17$\pm$0.06 & 1.9$\pm$0.3 & 2.6$\pm$0.2 \\
		HTM-A122	& AR	& 6	 & 1064 & 1.46 & 16.8 	&		& -1.13$\pm$0.08 & 1.9$\pm$0.3 & 2.5$\pm$0.2 \\
		HTM-A108	& HR	& 38 & 1064 & 0.56 & 6.5 	&		& -0.21$\pm$0.01 & 1.5$\pm$0.2 & 2.7$\pm$0.1 \\
		HTM-A125	& HR	& 38 & 1064 & 0.56 & 6.5 	&		& -0.18$\pm$0.01 & 1.5$\pm$0.2 & 2.7$\pm$0.1 \\	
	\end{tabular}
	\end{ruledtabular}
	\footnotetext[1]{samples with the same coating specifications have been coated together, by pairs.}
\end{table}
\endgroup
The results for coating stacks deposited on fused-silica cantilever blades are listed in Table \ref{TABstacksBlades}. On FIG.~\ref{FIGdeltaVSn} we show the excess loss,
\begin{equation}
\label{EQrelExcLoss}
\Delta = \langle \phi_m \rangle - \phi_e \ ,
\end{equation}
as a function of $N$. Indeed, FIG.~\ref{FIGdeltaVSn} clearly demonstrates that the measured loss is larger than the expectation, indicating that the loss of the coating stack is larger than the linear combination of the loss of the single layers. Then Eq.(\ref{EQsimpleMod}) is no longer verified and the total loss of the oscillator must account for an additional loss term.

\begin{figure}
	\includegraphics[width=0.75\textwidth]{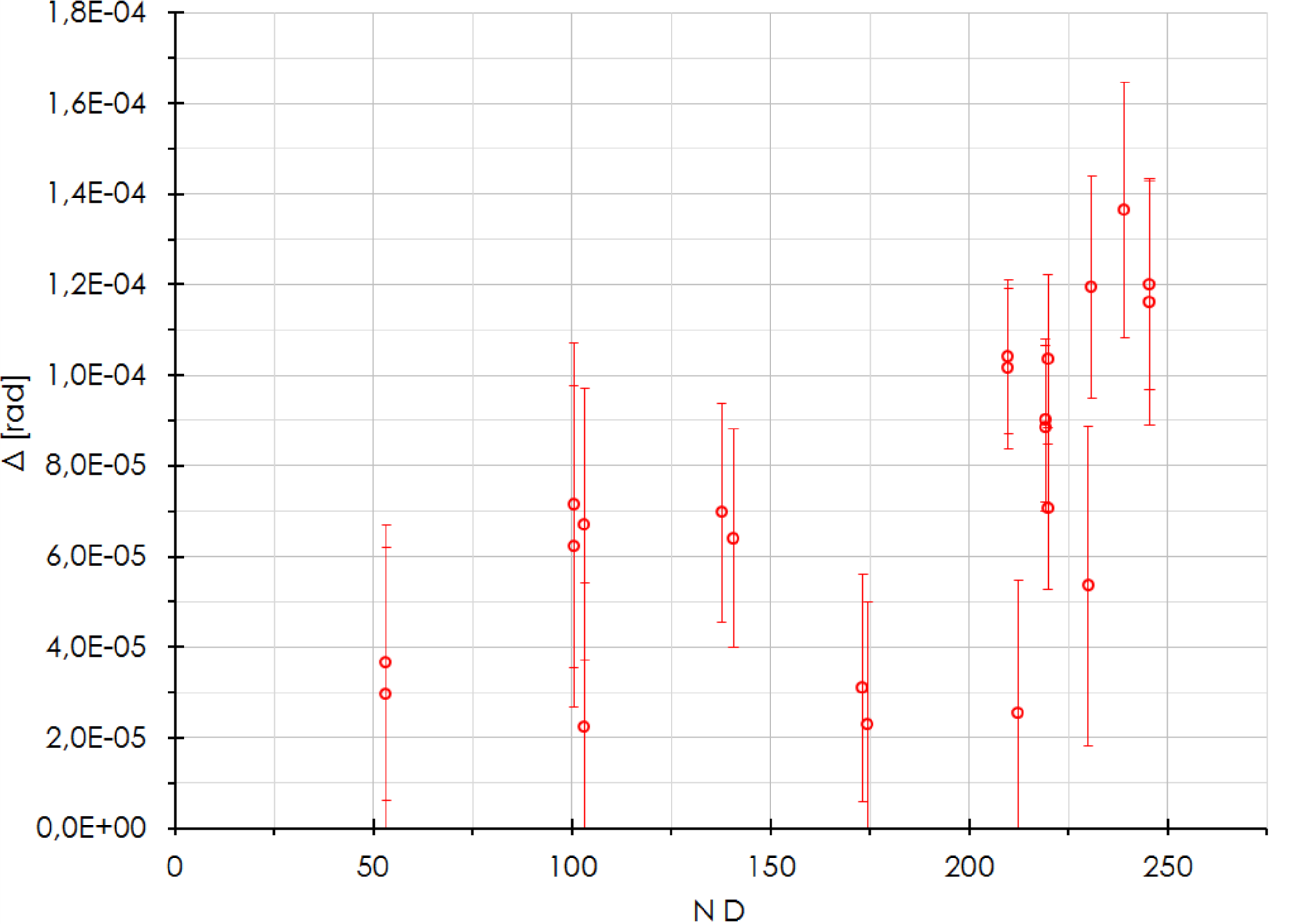}
	\caption{\label{FIGdeltaVSnd} Excess loss $\Delta$ of coating stacks deposited on cantilever blades, as a function of the product $N D$. Linear fitting of data yields a slope $\phi_i = 4 \pm 2 \times 10^{-7}$ in Eqs.(\ref{EQmeasPhiCmod}) and (\ref{EQphiE}). The large dispersion in the data, however, does not allow to confirm a linear dependence.}
\end{figure}

In order to test the existence of a loss at the interface between the layers, we draw in FIG.~\ref{FIGdeltaVSnd} a graph of $\Delta$ as a function of the product $ND$. If Eqs.(\ref{EQmeasPhiCmod}) and (\ref{EQphiE}) hold, we expect $\Delta$ to vary linearly and $\phi_i$ to be its slope. Unfortunately, the dispersion of data points in FIG.~\ref{FIGdeltaVSnd} does not allow to confirm or to rule out a linear dependence of $\Delta$. The linear fitting yields $\phi_i = 4 \pm 2 \times 10^{-7}$. At the same time, the dispersion might be the indication that an unknown phenomenon is taking place.

\begin{figure}
	\includegraphics[width=0.75\textwidth]{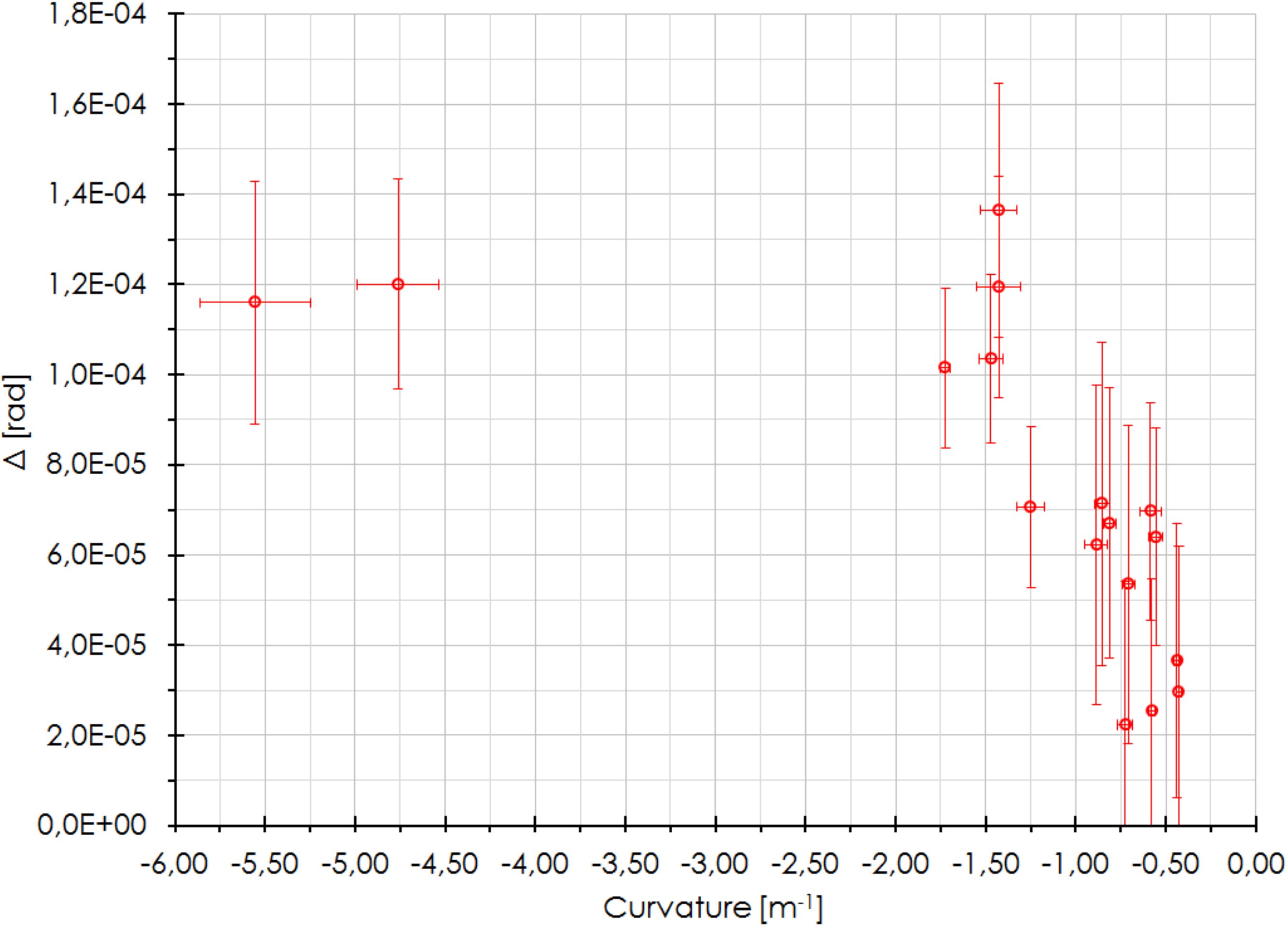}
	\caption{\label{FIGdeltaVScurv} Excess loss $\Delta$ of coating stacks deposited on cantilever blades, as a function of the curvature $1/R$. Excess loss is seen to grow exponentially with curvature.}
\end{figure}

In FIG.~\ref{FIGdeltaVScurv}, the excess loss $\Delta$ is plotted against the residual curvature $1/R$ after the post-coating annealing. All the samples showed residual compressive stress, denoted by a negative sign of the curvature. The graph shows that $\Delta$ is a function of the curvature, which in turn is determined by the internal stresses in the sample. In other words, the measured coating loss seems to be unexpectedly a function of $1/R$. Besides, even more surprisingly, the excess loss seems to increase exponentially with the curvature, i.e. $\Delta$ is much larger in bent samples where the coating is in principle more relaxed.

To further test the curvature effect, we deposited a nominally identical coating stack on the second surface of the cantilever blades H-5, H-6, HTM-A108 and HTM-A125. The curvature of samples H-5 and H-6 had not been measured after the deposition of the first stack, but their bending was large enough to be appreciated by eye. After the deposition of the second stack, their RoC was so large that it could not be measured with our apparatus. The second stack deposited on samples HTM-A108 and HTM-A125 turned out to be 5\% thicker than the first one, so that the curvature did not decrease to zero as expected. In any case, for three samples out of four (H-5, HTM-A108 and HTM-A125) the deposition of the second stack noticeably decreased the measured coating loss $\langle \phi_{m} \rangle$. For H-6, the loss decrease is comparable to the experimental uncertainty. These results, reported in Table \ref{TAB2stacksBlades}, confirm the dependence of the measured coating loss on the stress within the sample.
\begingroup
\squeezetable
\begin{table}
	\caption{\label{TAB2stacksBlades} Specifications and results of fused-silica cantilever blades before and after the deposition of a second coating stack: sample, coating design, number of coating layers $N$, thickness ratio $r$, longitudinal mode order $k$, total loss angle $\phi_k^{cs}$, simulated dilution factor $D$, expected coating loss $\phi_{e}$, measured coating loss $\langle \phi_{m} \rangle$, RoC $R$. Loss angles are given in units of $10^{-4}$ rad.}
	\begin{ruledtabular}
	\begin{tabular}{l c c c c c c c c c c}
		sample\footnotemark[1] & coating & $N$ & $r$ & $k$ & $\phi_k^{cs}$ & $D$ & $\phi_{e}$ & $\langle \phi_{m} \rangle$ & $R$ [m] \\ \hline
		H-5	& HR & 18 & 0.35 & 0 & 0.186 & 12.2 & 1.3$\pm$0.2 & 2.2$\pm$0.1 & \\
			& & & & 1 & 0.190 & & & & \\ \hline
		H-5	& 2 $\times$ HR\footnotemark[2] & 18/18 & 0.34 & 0 & 0.281 & 6.3 & 1.2$\pm$0.2 & 1.7$\pm$0.1 & $\infty$ \\
			& & & & 1 & 0.288 & & & & \\
			& & & & 4 & 0.291 & & & & \\ \hline
		H-6	& HR & 18 & 0.35 & 0 & 0.187 & 12.2 & 1.3$\pm$0.2 & 2.1$\pm$0.1 & \\
			& & & & 1 & 0.185 & & & & \\
			& & & & 2 & 0.187 & & & & \\ \hline
		H-6	& 2 $\times$ HR\footnotemark[2] & 18/18 & 0.34 & 0 & 0.310 & 6.3 & 1.2$\pm$0.2 & 1.9$\pm$0.1 & $\infty$ \\
			& & & & 1 & 0.336 & & & & \\
			& & & & 2 & 0.306 & & & & \\
			& & & & 3 & 0.326 & & & & \\
			& & & & 4 & 0.337 & & & & \\ \hline
		HTM-A108	& HR & 38 & 0.56 & 0 & 0.443 & 6.5 & 1.5$\pm$0.2 & 2.7$\pm$0.1 & -0.21$\pm$0.01 \\
			& & & & 1 & 0.449 & & & & \\
			& & & & 2 & 0.435 & & & & \\ \hline
		HTM-A108	& 2 $\times$ HR\footnotemark[2] & 38/38 & 0.56 & 0 & 0.726 & 3.2 & 1.5$\pm$0.2 & 2.3$\pm$0.1 & -0.30$\pm$0.01 \\
			& & & & 1 & 0.745 & & & & \\
			& & & & 2 & 0.756 & & & & \\
			& & & & 3 & 0.754 & & & & \\
			& & & & 4 & 0.778 & & & & \\ \hline
		HTM-A125	& HR & 38 & 0.56 & 0 & 0.441 & 6.5 & 1.5$\pm$0.2 & 2.7$\pm$0.1 & -0.18$\pm$0.01 \\
			& & & & 1 & 0.444 & & & & \\
			& & & & 2 & 0.401 & & & & \\
			& & & & 3 & 0.467 & & & & \\
			& & & & 4 & 0.457 & & & & \\ \hline
		HTM-A125	& 2 $\times$ HR\footnotemark[2] & 38/38 & 0.56 & 0 & 0.734 & 3.2 & 1.5$\pm$0.2 & 2.3$\pm$0.1 & -0.26$\pm$0.01 \\
			& & & & 1 & 0.747 & & & & \\
			& & & & 2 & 0.732 & & & & \\
			& & & & 3 & 0.768 & & & & \\
			& & & & 4 & 0.763 & & & & \\
	\end{tabular}
	\end{ruledtabular}
	\footnotetext[1]{samples with the same coating specifications have been coated together, by pairs.}
	\footnotetext[2]{both surfaces of the sample have been coated.}
\end{table}
\endgroup

We also deposited the same two coating stacks (one with $N = 18$ with $r = 0.35$, the other having $N = 38$ with $r = 0.56$) on thicker fused-silica disks, to measure the coating loss in a condition of maximum stress. Because of the larger thickness, we assumed that the curvature of these substrates did not change appreciably after coating. As shown in Table \ref{TABstacksDisks}, the excess loss is slightly lower in disks than in cantilever blades. This is a further confirmation of the stress effect on the measured coating loss.
\begingroup
\squeezetable
\begin{table}
	\caption{\label{TABstacksDisks} Specifications and results of fused-silica disks coated with stacks: sample, coating design, number of coating layers $N$, thickness ratio $r$, mode order\footnote{$(r, a)_k$ is the $k$-th mode with $r$ radial and $a$ azimuthal nodes.} $(r, a)_k$, simulated dilution factor $D_k$, expected coating loss $\phi_{e}$, measured coating loss $\langle \phi_{m} \rangle$. Loss angles are given in units of $10^{-4}$ rad.}
	\begin{ruledtabular}
	\begin{tabular}{l c c c l c c c c c}
		sample & coating & $N$ & $r$ & $(r, a)_k$ & $D_k$ & $\phi_{e}$ & $\langle \phi_{m} \rangle$ \\ \hline
		39	& HR & 18 & 0.32 & (0, 2)$_1$ & 237.1 & 1.2$\pm$0.2 & 1.5$\pm$0.1 \\
				& & & & (0, 3)$_4$ & 237.6 & & \\
				& & & & (0, 5)$_{10}$ & 240.3 & & \\
				& & & & (1, 2)$_{12}$ & 236.0 & & \\
				& & & & (0, 6)$_{15}$ & 242.2 & & \\ \hline
		38	& HR & 38 & 0.56 & (0, 2)$_1$ & 112.5 & 1.5$\pm$0.2 & 2.1$\pm$0.1 \\
				& & & & (0, 3)$_4$ & 112.6 & & \\
				& & & & (0, 5)$_{10}$ & 113.7 & & \\
				& & & & (1, 2)$_{12}$ & 111.1 & & \\
				& & & & (0, 6)$_{15}$ & 114.6 & & \\
	\end{tabular}
	\end{ruledtabular}
\end{table}
\endgroup

In summary, by looking at the results of Tables \ref{TAB2stacksBlades} and \ref{TABstacksDisks}, it is clear that the residual stress is correlated with the measured excess loss. However, our attempts to build up a relatively simple model of the excess loss (like $\Delta = \alpha/R + ND\phi_i$ or $\Delta = A e^{\alpha/R} + ND\phi_i$) proved to be inconclusive because of the poor quality of the fitting, yielding very large uncertainties on the parameters $\alpha$ and $\phi_i$. Certainly, the loss excess does not have a trivial explanation. Its unexpected complexity calls for further tests and deeper analyses which seems to exceed the boundaries of this present work.

\subsubsection{Interfaces}
\begin{figure}
	\includegraphics[width=1.00\textwidth]{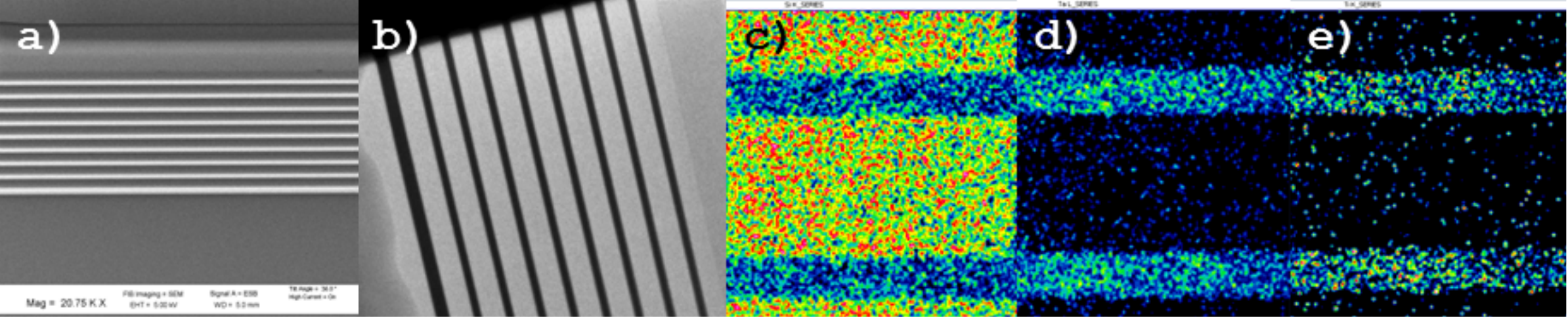}
	\caption{\label{FIGclymTEM} Analyses of a coating stack: a) SEM image (bright: TiO$_2$Ta$_2$O$_5$ layers); b) bright-field TEM image (dark: TiO$_2$Ta$_2$O$_5$ layers); c)-e) from left to right, EDX cartography of the confinement of Si, Ta and Ti within 5 layers. Courtesy of the CLYM.}
\end{figure}

We used the FIB sample from a coating stack with $N = 18$ and $r = 0.35$ for the analyses. The SEM and TEM imagery revealed a regular structure with no appreciable defect. According to the images, the thickness of the coating blade penetrated by the electron-beam probe is $40 \pm 10$ nm. An EDX cartography showed a sharp transition between the layers, once the background noise neglected. Some exemplary pictures are shown on FIG.~\ref{FIGclymTEM}.

\begin{figure}
 \subfigure[]{
	\includegraphics[width=0.75\textwidth]{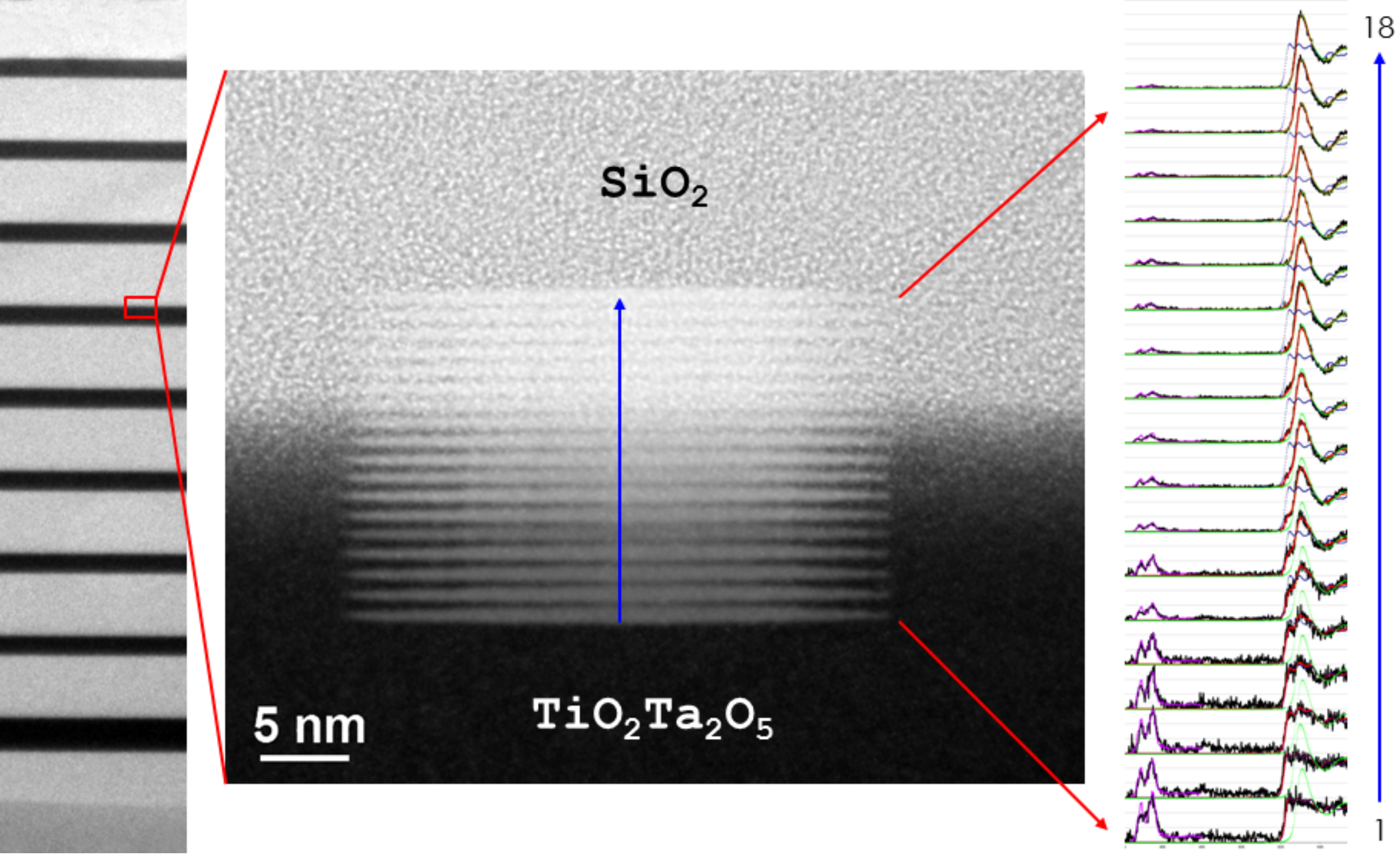}
 }
 \subfigure[]{
   \includegraphics[width=0.75\textwidth]{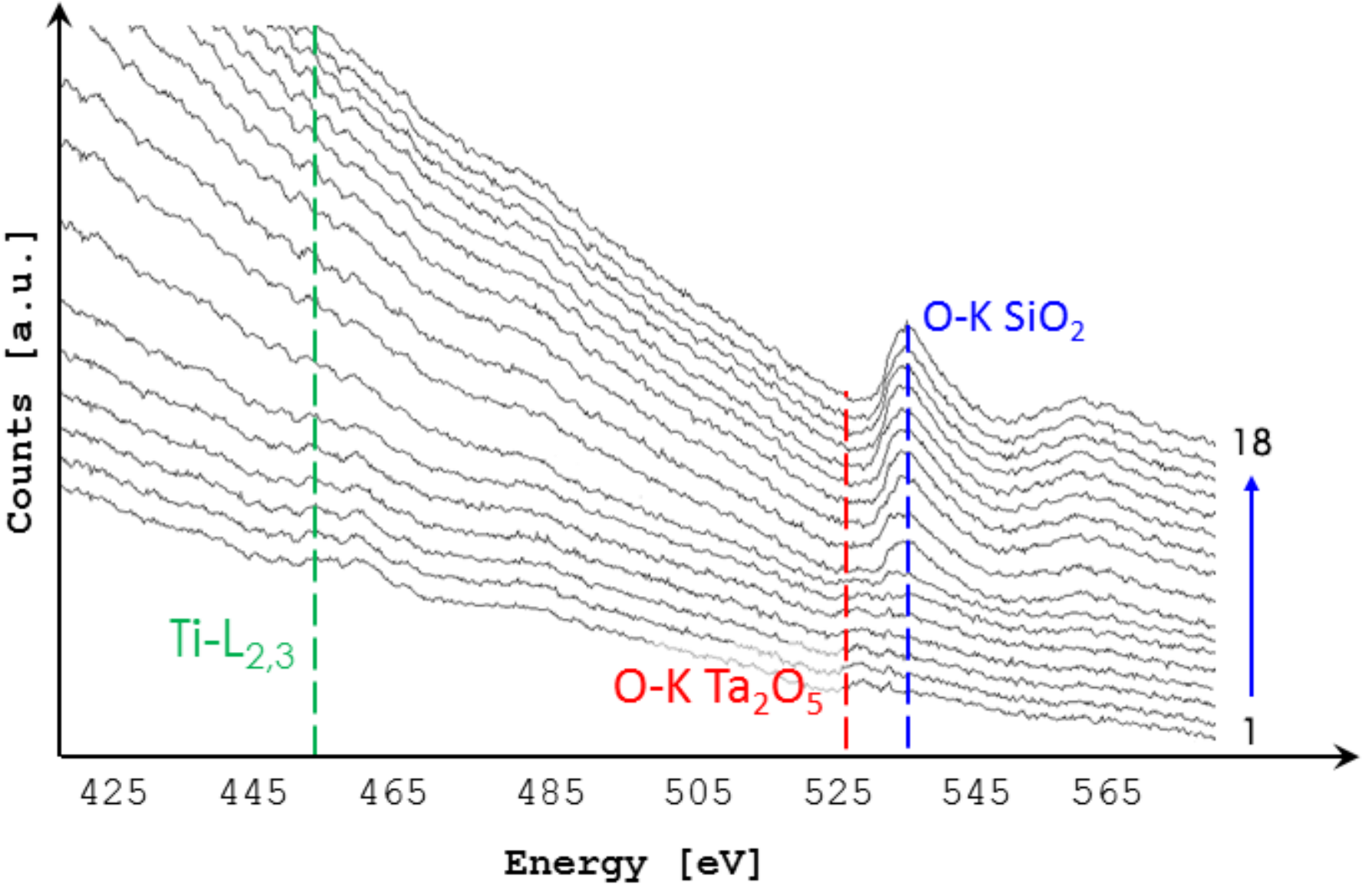}
 }
	\caption{(Color online) TEM images and EELS line scans at the interface between doped tantala (dark) and silica (bright), in a stack with with $N = 18$ and $r = 0.35$: a) from left to right, location of the interface to be analyzed within the coating stack (transition between the fifth and the sixth doublet), zoom on the interface with the trace of eighteen horizontal line scans (the blue arrow shows the sequence of the scans), line scan and fitting spectra (backgrounds subtracted); b) raw EELS line scan spectra (with background noise) and atomic-transition energies (dashed lines) used in the quantitative analysis. Courtesy of the CLYM.}
	\label{FIGeelsScans}
\end{figure}

For a more detailed analysis of the interface, we used the EELS data collected from eighteen consecutive line scans at the edge of two layers, as shown on FIG.~\ref{FIGeelsScans}. The line scans have been performed with an electron-beam probe of nominal width of 1 nm, separated by 1 nm steps along the direction perpendicular to the layer thickness. The acquired signal was integrated to yield a single spectrum per line scan. Two additional reference spectra have been acquired in the middle of a silica and of a titania-doped tantala layer, respectively. In these spectra, we took the K edge of oxygen (K-O) in SiO$_2$ and in Ta$_2$O$_5$ for quantification in the next analysis\footnote{More details on the atomic transitions used in EELS may be found elsewhere \cite{Egerton}.}. An additional spectrum for the Ti-L$_{2,3}$ line has been acquired separately on a rutile powder, to have a reference for the TiO$_2$ signature in the spectrum of the titania-doped tantala layer. Finally, a linear combination of the three reference spectra was fitted to each line scan to quantify the content of SiO$_2$ and Ta$_2$O$_5$ at the interface. All the spectra used in the analysis (raw line scans, line scans with background noise subtracted, fitting spectra) are shown on FIG.~\ref{FIGeelsScans}.
\begin{figure}
	\includegraphics[width=0.75\textwidth]{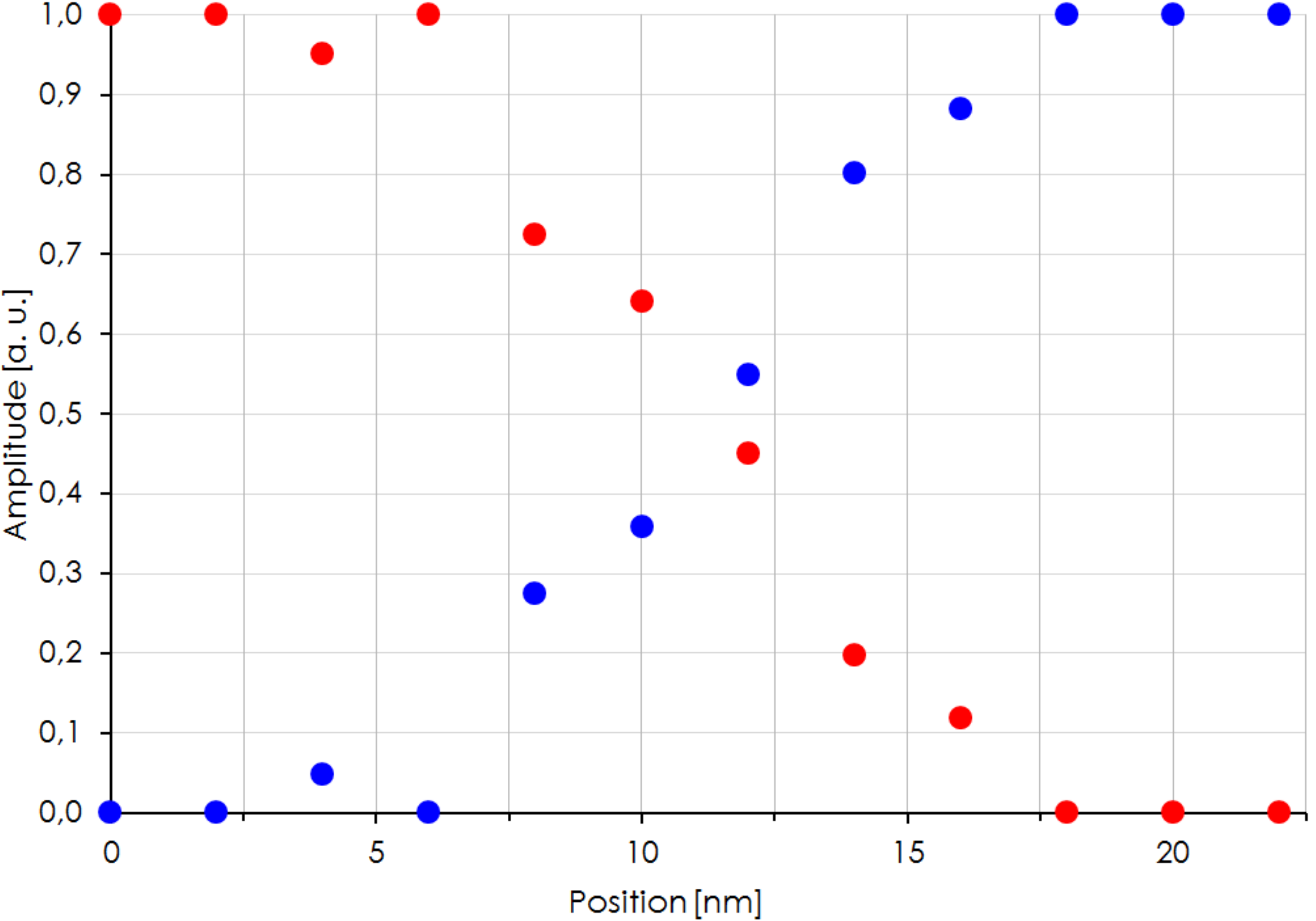}
	\caption{\label{FIGeelsConcentr} (Color online) Normalized atomic ratios (from fitting coefficients of K-O edges in EELS spectra, in arbitrary units) of SiO$_2$ (blue) and Ta$_2$O$_5$ (red), as a function of the position across an interface in a stack with $N = 18$ and $r = 0.35$.}
\end{figure}

The final results of the EELS analysis is presented on FIG.~\ref{FIGeelsConcentr}. Here, the normalized amplitude of the fitting coefficients of K-O edges in SiO$_2$ and Ta$_2$O$_5$ are shown as a function of the position in the interface (the redundant information from the TiO$_2$ coefficients is neglected). The graph indicates clearly that there is a gradient of species concentration at the interface between layers. Looking at the spatial width of the gradient, the interface appears to be $10 \pm 1$ nm thick. Once corrected from the systematic enlargement experienced by the beam in penetrating the sample, the actual estimated thickness of the interface is reduced to $5 \pm 1$ nm at most. If our analysis is correct, this value is rather large. It would correspond to a few percent of each layer thickness, yielding (for $N>>1$) a consistent fraction of mixed material in the coating. However, this value is compatible with a tilt of the coating section of about 8$^{\circ}$ with respect to the impinging beam, so the accuracy of our measurements appears to be limited by the intrinsic configuration of the experimental setup. Further measurements will be needed to address the issue of the interface thickness definitively.

\section{Discussion}
\subsection{Summary}
Our findings indicate that the loss of mono-layers is independent of the coating thickness and of the stress of the sample. Furthermore, the loss from the interface with the substrate is eventually negligible. The average measured loss of silica mono-layers is $\langle \phi_m \rangle_{\text{SiO}_2} = 4.5 \pm 0.3 \times 10^{-5}$, in agreement with the outcome of independent measurements. The average measured loss of titania-doped tantala mono-layers is $\langle \phi_m \rangle_{\text{TiO}_2\text{Ta}_2\text{O}_5} = 2.4 \pm 0.3 \times 10^{-4}$, which, for reasons yet to be understood, is in contrast with the result from independent experiments.

The synthesis concerning our results from stacks is far more complex. We measured several thin cantilever blades, with a single or a double coating, and a few thicker disks with a single coating. A clear outcome is that the measured loss is larger than expected (provided that the expectation is correct, of course), though the cause of this loss excess is still not known. Double-coated cantilever blades and disks show similar results, but the loss excess is smallest when measured on disks.

In the attempt to understand the origin of this phenomenon, we looked first for the presence of an unknown dissipative mechanism at the interface between layers. Then, pushed by some new unexpected evidence from our results, we tested also the hypothesis of a stress-dependent loss. Unfortunately, we could not conclusively prove any of these speculations. Our TEM analyses could not determine unequivocally the steepness of the interface between the layers, though we know now that it should be 5 nm wide at the most. If any, the interface loss would be only a tiny fraction (between 1{\textperthousand} and 1\%, depending on the sample) of the observed excess loss. On the other hand, we could not find any simple model to satisfactorily describe the dependence of the excess loss on stress. Regrettably, it seems that we cannot proceed deeper in our analysis with the data presently in our possession.

\subsection{Conclusions}
In this work we presented the results of several years of coating characterization carried out at LMA. This database comprehends results from a large variety of titania-doped tantala and silica coatings of different thickness, number of layers and thickness ratios. The coatings have been characterized on fused-silica substrates of different geometries on different suspension systems.

The results for mono-layer coatings appear well assessed, globally. The measured loss of each material is constant with respect to the specificity of the samples (thickness, residual stress), as one would expect for an intrinsic loss mechanism in the material.

For coating stacks, we could not unequivocally prove the existence of a lossy interface of mixed materials between the layers. Nevertheless, our data demonstrate the existence of an excess loss which was totally unexpected from a linear combination of measured mono-layer loss. Moreover, this excess loss appears to be related to the residual stress inside the samples. Some of the coatings presented here have the same design as those deposited on the input ($N = 18$, $r = 0.35$) and the output ($N = 38$, $r = 0.56$) test masses of Advanced LIGO and Advanced Virgo, so our measurements -- which are their most up-to-date characterization -- have a particular relevance for the noise budget of those detectors.

Though the origin of the excess loss still remains unknown, it is possible to draw some conclusions from our analysis. From now on, the stress inside the coating should be considered as a relevant parameter for the characterization of mechanical loss. According to our results, the measured loss is closest to the expectation if the coating stress is not relaxed, on relatively stiff substrates. This configuration surely does have the advantage of being closer to the actual operation condition of coatings in interferometric gravitational-wave detectors, which currently use very thick (20 cm) substrates. Of course, for coating characterization, the substrate thickness should be determined as a trade-off between the need to keep the coating stressed and the requirement of minimizing the dilution factor to enhance the effect of the coating loss.

Finally, we hope that this work might help in developing a new protocol of mechanical-loss characterization and possibly trigger new studies on the unsolved issues we raised.

\section{Acknowledgments}
The authors acknowledge T. Epicier and B. Van De Moortele of the CLYM for the electron microscope analyses, and are grateful to the LABEX Lyon Institute of Origins (Grant No. ANR-10-LABX-0066) of the Universit\'e de Lyon for its financial support within the program ``Investissements d'Avenir'' (Grant No. ANR-11-IDEX-0007) of the French government operated by the National Research Agency (ANR).

\end{document}